\documentclass[ review]{elsarticle}
\usepackage{setspace}
\usepackage{enumerate}
\usepackage{graphicx}
\usepackage{amsmath, bm}
\usepackage{float}
\usepackage{mdframed}
\usepackage{multirow}
\usepackage{textcomp}
\usepackage{subcaption}
\usepackage{fixltx2e}
\usepackage{url}
\usepackage{makecell}
\usepackage{amssymb}
\usepackage{bm}
\usepackage{textcmds}
\usepackage{hyperref}
\biboptions{numbers,sort&compress}

\usepackage{tikz}
\usetikzlibrary{shapes.geometric, arrows}
\usepackage{caption}  % For customizing figure captions

\def\be{\begin{equation}}\def\ee{\end{equation}}
\def\ba{\begin{array}}\def\ea{\end{array}}
\def\bfg{\begin{figure}}\def\efg{\end{figure}}
\makeatletter
\def\fps@figure{htbp}
\makeatother
\usepackage{stackengine}
\stackMath
\newcommand\tenq[2][1]{%
 \def\useanchorwidth{T}%
  \ifnum#1>1%
    \stackunder[0pt]{\tenq[\numexpr#1-1\relax]{#2}}{\scriptscriptstyle\sim}%
  \else%
    \stackunder[1pt]{#2}{\scriptscriptstyle\sim}%
  \fi%
}

\RequirePackage[margin=20mm]{geometry}

\journal{Computers \& Mathematics with Applications}
\begin{document}

\begin{frontmatter}

% \title{Mathematical analysis of elastic wave propagation in initially stressed and rotating functionally graded viscoelastic piezoelectric materials model with interfacial crack}
% \title{Wave Propagation Analysis in Functionally Graded Piezoelectric-Viscoelastic Media with Interface Defects}
\title{Integral transform technique for determining stress intensity factor in wave propagation through functionally graded piezoelectric-viscoelastic structure}

\author[label1]{Diksha}
\author[label1]{Soniya Chaudhary*}
\author[label1]{Pawan Kumar Sharma}

\cortext[cor1]{Corresponding author: soniyachaudhary18@gmail.com}
\address[label1]{Department of Mathematics and Scientific Computing, National Institute of Technology Hamirpur, Himachal Pradesh, 177005, India}

\begin{abstract}
This study employs an integral transform approach for Love wave propagation in a rotating composite structure having an interfacial crack. The structure comprises an initially stressed functionally graded piezoelectric viscoelastic half-space bonded to a piezoelectric viscoelastic half-space. The study focuses on two material systems: Epoxy-BNKLBT paired with Epoxy-KNLNTS and Epoxy-BNKLBT paired with Epoxy-PZT7A. The viscoelastic materials are modeled to reflect their complex behavior under rotational and stress conditions. The Galilean transformation is applied to convert the Cartesian coordinates system into a moving reference frame aligned with the Love wave's propagation. Employing Bessel function properties, the system is converted into a set of double integral equations and subsequently reformulated into simultaneous Fredholm integral equations. Numerical solutions to these Fredholm integral equations are used to calculate the electric displacement intensity factor (EDIF) and stress intensity factor (SIF) near the interfacial crack.
The key objective of this study is to visualize the impact of different material parameters, like piezoelectric constants, dielectric constants, initial stress, interface electric displacement, interface stress, and rotation, on SIF and EDIF.
The investigations of this study will be helpful for advanced technologies like surface acoustic wave (SAW) sensors and piezoelectric actuators, as well as to enhance SAW bio-sensor sensitivity and stability for early cancer detection and biomedical implants. 
\end{abstract}
\end{frontmatter}
\section{Introduction}
Acoustic wave-based devices have gained significant attention in scientific research. Since their development, they have been widely applied in engineering instruments such as oscillators, resonators, sensors, and amplifiers, and have also shown promise in medical applications like cancer detection and therapy monitoring. Among these, Love wave sensors stand out for their high sensitivity and low loss in fluid environments, making them ideal for measuring liquid properties and detecting chemical warfare agents \cite{wang1997cylindrical, royer1999elastic, du1996study,pedrick2007sensitivity, moreira2008theoretical,ramshani2015sh}. With the rising prevalence of cancer, SAW bio-sensors have also become crucial for the early detection of tumor metastasis by identifying circulating tumor cells. Recent research has focused on optimizing bio-sensors using interdigital transducers (IDTs) to enhance sensitivity and temperature stability \cite{fall2018optimization, lo2020frequency,alam2024introducing}.

The incorporation of thin films, whether piezoelectric or non-piezoelectric, to SAW devices, greatly affects their properties. Common piezoelectric materials include ZnO, PZT, PLZT, and  AIN, while PMMA and silicone are often used as non-piezoelectric films. Piezoelectric materials, commonly used in engineering and technological applications, are particularly susceptible to reduced service life under long-term periodic loads. The classical theory of elasticity faces challenges in addressing wave propagation on micro and nano scales due to its inability to account for a characteristic length scale. Mechanical failure of piezoelectric ceramic layers remains a significant concern in practical structures. To enhance the lifetime and reliability of these materials, the method of incorporating functionally graded materials is applied to piezoelectric materials, supported by advancements in modern material processing technology \cite{hu2005anti,feng2006dynamic,zhou2007investigation,li2009interaction,lu2021characteristic}. In engineering, viscoelastic interfaces may be deliberately incorporated to impart attenuation and energy dissipation properties to inherently rigid piezoelectric devices \cite{hashin1991composite,li2009fracture,ngak2018dynamic}. Therefore, incorporating viscoelastic layers is crucial when designing and applying multi-layer piezoelectric components. Studies have shown that these layered structures enhance mass sensitivity in liquid media without compromising frequency stability or Q-factor \cite{han2014mass, singh2023propagation}. 
The characteristics of Love waves in such multi-layered composites, particularly when bonded to elastic substrates, have been extensively analyzed theoretically. These waves are particularly useful in practical applications such as gas and liquid detection due to their superior performance compared to Rayleigh waves in these environments \cite{sayago2016graphene,li2009interaction,singhal2018liouville}. 

The transmission of elastic waves through functionally graded piezoelectric material  (FGPM) structures which have been made of different materials has been thoroughly studied by various researchers \cite{liu2001effect,danoyan2007surface, du2008propagation,eskandari2008love}. Li \textit{et al.} \cite{li2004love}, and Qian \textit{et al.}
\cite{qian2007transverse} examined the behavior of propagation of Love wave in various types of FGPM layered structures. Layered structures, when carefully designed with appropriate boundary conditions, can further improve the performance of SAW devices. For instance, Love wave sensors have proven effective in detecting bacteria, cancer cells, and DNA hybridization, showcasing their notable benefits in biological applications \cite{hur2005development, chang2014label, borodina2018sensor}. This body of work underscores the importance of continued research into the material and structural optimization of SAW devices to expand their real-world applications.

During the production of SAW (surface acoustic wave) devices, micro-cracks at the interface are a significant concern, as these defects can lead to stiffness reduction, delamination, and eventually structural failure under operational loads. The study of wave interaction with interfacial defects, particularly interfacial cracks, has been a focal point of research since the 1960s. Williams \cite{williams1957stress} employed the complex variable technique to investigate the stress and displacement distributions near crack tips. Erdogan \cite{erdogan1965stress} advanced this methodology to accommodate additional Griffith cracks. Rice and Sih \cite{rice1965plane} employed the R-integral technique to address the issue of cracks in confined, heterogeneous materials. Willis \cite{willis1971fracture} formulated specific design principles in his research focusing on interfacial cracks. Dwivedi \textit{et al.} \cite{dwivedi1980pair, dwivedi1981stress} explored the challenges associated with cracks in various types of media. Researchers like Krenk \textit{et al.} \cite{krenk1982elastic}, Keer \textit{et al.} \cite{keer1984resonance}, and Yang \textit{et al.} \cite{yang1985elastic} have developed various analytical methods, such as Green's functions, integral equation methods, and dislocation density functions, to analyze the propagation of elastic waves emanating from cracks and to accurately determine the SIF associated with these cracks. These studies highlight the critical need to understand and mitigate the effects of interfacial defects to improve the reliability and longevity of SAW devices.
Building on previous work, Wei \textit{et al.} \cite{wei2002scattering} extensively studied how a non-uniform plane surface horizontal wave scatters, refracts, and reflects at an interfacial crack between two distinct heterogeneous viscoelastic materials. Hu \textit{et al.} \cite{hu2005anti} examined a crack in a FGPM interlayer positioned between two different piezoelectric half-planes, considering the effects of anti-plane mechanical loading and in-plane electric fields.
Ueda \cite{ueda2006finite} investigated a FGPM strip with a crack under the effects of in-plane electric loading. In a separate study \cite{ueda2007electromechanical}, Ueda analyzed a FGPM layer with a parallel crack subjected to both in-plane mechanical and electric forces.
To analyze the effects of elastic waves on fractures in a multi-layered piezoelectric composite structure, the complex boundary integral equation method was utilized \cite{singh2009scattering, fomenko2021advanced,zhang2024sh, singh2024scattering}.
 Studies related to the presence of cracks or punches on wave propagation have been conducted and documented in various works, including \cite{afshar2018several, ozdemir2020dynamic,yazdi2021efficient,awasthi2022griffith, solovyev2023numerical,boroujerdi2024multiple}. Numerous distinguished researchers have devoted significant efforts to addressing the challenges of crack propagation, leading to the formulation of methods for determining the SIF under the influence of wave propagation \cite{das1998stress,kim2022extraction,krpensky2023exact, kumar2024moving}.
The presence of interfacial micro-cracks significantly threatens the structural integrity of engineering materials. Understanding wave interactions with these defects is crucial for enhancing material durability and preventing failures. Despite progress in studying these interactions, a more detailed understanding of wave propagation in composite structures is needed. Accurately predicting SIFs in multi-layered structures remains challenging due to the limitations of simpler models. The literature review reveals a notable research gap: the propagation of Love waves from a crack at the interface in an initially stressed functionally graded rotating piezoelectric viscoelastic material half-space coupled with a rotating piezoelectric viscoelastic material half-space has not been previously explored. 

This study addresses this gap by examining the interactions of Love waves with interfacial cracks in composite structures. Specifically, it investigates two material systems: one consisting of Epoxy-BNKLBT and Epoxy-KNLNTS and the other of Epoxy-BNKLBT and Epoxy-PZT7A. The focus is on the behavior of Love waves emanating from a crack at the interface within these bimaterials, providing new insights into wave propagation in these advanced composite structures. 
The paper is organized into the following sections: Section \ref{theoretical statement} covers the analytical framework, which includes subsection \ref{problem statement} detailing the problem statement, subsection \ref{Formulation of the problem} presenting the problem formulation and governing equations, including the Galilean transformation to the moving reference frame, subsection \ref{Solution of the field variables} describing the solution of the field variables, subsection \ref{section bc} addressing the boundary conditions, subsection \ref{Formulation of Stress Intensity and Electric Displacement Intensity Factors} deriving the final expressions for the SIF and EDIF.  Section \ref{Numerical Calculation and Graphical Representation} presents the numerical calculations and graphical representations. Finally, Section \ref{conclusion} summarizes the key findings and insights from the numerical and comparative analysis, concluding the study.
\section{Analytical framework}
\label{theoretical statement}
\subsection{Problem definition}
\label{problem statement}

\begin{figure}
    \centering
    \includegraphics[width=1\linewidth]{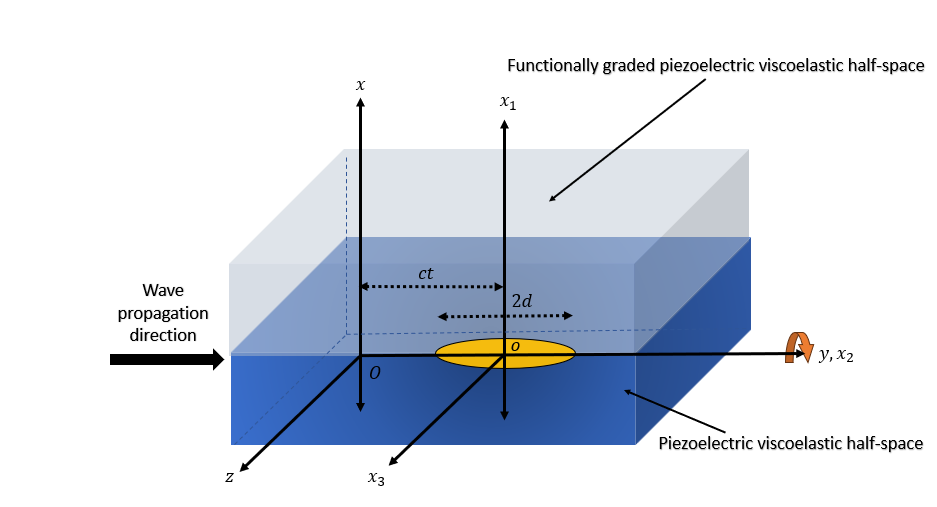}
    \caption{Schematic representation of the physical problem.}
    \label{fig1}
\end{figure}

Figure \ref{fig1} depicts the arrangement used to examine the propagation of surface waves at a spatial frequency, $\omega$, in a rotating medium composed of piezoelectric and viscoelastic half-spaces. The system comprises two main regions: an upper half-space, which is initially stressed functionally graded piezoelectric-viscoelastic material (FGPVM) half-space ($x \geq 0$), and a lower half-space, which is a homogeneous piezoelectric-viscoelastic material (PVM) half-space ($x \leq 0$). The upper layer features material properties such as elasticity and piezoelectric coefficients that vary gradually along its thickness. In contrast, the lower layer has consistent piezoelectric-viscoelastic properties throughout and lies directly beneath the functionally graded half-space.
Key elements in the figure include the wave propagation direction, indicated by an arrow pointing horizontally to the left, showing the path of waves through the medium. The coordinate axes are defined such that the $x$-axis represents depth within the layers, the $y$-axis runs horizontally and aligned with the wave propagation direction, and the $z$-axis extends perpendicularly to the plane. A yellow elliptical region appears at the interface between the two half-spaces, representing a circular defect where wave interactions are of particular interest. This defect’s position, denoted by $o$, is situated at the boundary between both half-spaces. The distance $2d$ signifies the characteristic dimension across this defect, typically associated with its width, which plays a critical role in this study. Consider \(\mathbf{u} = \left(u^{(p)}, \: v^{(p)}, \: w^{(p)}\right)\) as the components of displacement in their respective directions (i.e. \(x\), \(y\), and \(z\) directions), respectively. Given that a Love wave propagates along the \(y\)-axis and induces particle displacement exclusively in the \(z\)-axis, therefore:
\begin{equation}
    u^{(p)}=0; \:\:\: v^{(p)}=0; \:\:\: w^{(p)}=w^{(p)}(x,\:y,\:t);\:\:\: \frac{\partial}{\partial z}=0.
    \label{displacement_component}
\end{equation}

\subsection{ Mathematical model and problem formulation}
\label{Formulation of the problem}
The general form of the governing equation of motion and Gauss's law for initially stressed piezoelectric-viscoelastic materials with a uniform angular velocity \(\mathbf{\Re^{(p)}} = (\Re^{(p)}_x, \: \Re^{(p)}_y, \: \Re^{(p)}_z)\) as outlined by Wang and Shang \cite{wang1997cylindrical}, is:
 \begin{equation}
\nabla \cdot \mathbf{T}^{(p)} + \nabla \cdot \left( \nabla \mathbf{u}^{(p)} \cdot \Im^{(p)} \right)  = \mathbf{F},\:\:\:
 \nabla \cdot \mathbf{D}^{(p)} + \nabla \cdot \left( \mathbf{u} \cdot \mathbf{D}^{0(p)} \right) = 0,
 \label{Equation for piezoelectric layer}
 \end{equation}
 where \(\mathbf{F}\) is the force vector given by:
 \begin{equation}
 \mathbf{F} = \rho^{(p)} \left[ \ddot{\mathbf{u}}^{(p)} + \left( \mathbf{\Re^{(p)}} \times \left( \mathbf{\Re^{(p)}} \times \mathbf{u}^{(p)}  \right) \right) + \left( 2 \mathbf{\Re^{(p)}} \times \dot{\mathbf{u}}^{(p)} \right) \right].
 \end{equation}
 The term $\nabla \cdot \mathbf{T}^{(p)}$ represents the divergence of the stress tensor $\mathbf{{T}^{(p)}}$, $ \mathbf{u}^{(p)}$ is mechanical displacement vector, $\Im^{(p)}$ represents the initial stress tensor, $\mathbf{{D}^{(p)}}$ is the electric displacement vector, $\mathbf{D}^{0(p)}$ represents the initial dielectric displacement vector, and $\rho^{(p)}$ is the mass density of the medium. The expression $\left( \mathbf{\Re^{(p)}} \times \left( \mathbf{\Re^{(p)}} \times \mathbf{u}^{(p)}  \right) \right)$ is the centripetal acceleration due to the varying time motion only, and $\left( 2 \mathbf{\Re^{(p)}} \times \dot{\mathbf{u}}^{(p)} \right)$ is the Coriolis acceleration. \\
%%%%%%%%%%%% MATRIX FORM-------------
The general three-dimensional constitutive relation for a piezoelectric-viscoelastic material can be expressed as \cite{royer1999elastic}:
\be
\mathbf{T^{'}}^{(p)} = \mathbf{C}^{(p)} \mathbf{S}^{(p)},
\label{constitutive relation}
\ee
where, \(\mathbf{T^{'}}^{(p)}\) is the vector of stress components and electric displacement components and \(\mathbf{S}^{(p)}\) is the vector of strain components and electric field components which are defined as:
\be
\begin{aligned}
\mathbf{T'}^{(p)} &= 
\begin{bmatrix}
T_{xx}^{(p)} & T_{yy}^{(p)} & T_{zz}^{(p)} & T_{yz}^{(p)} & T_{xz}^{(p)} & T_{xy}^{(p)} & D_x^{(p)} & D_y^{(p)} & D_z^{(p)}
\end{bmatrix} ^ \text{T}, \\
\mathbf{S}^{(p)} &= 
\begin{bmatrix}
S_{xx}^{(p)} & S_{yy}^{(p)} & S_{zz}^{(p)} & S_{yz}^{(p)} & S_{xz}^{(p)} & S_{xy}^{(p)} & E_x^{(p)} & E_y^{(p)} & E_z^{(p)}
\end{bmatrix}^\text{T},
\label{T(p)S(p)}
\end{aligned}
\ee
and \(\mathbf{C}^{(p)}\) is the stiffness matrix, which characterizes the material's response to mechanical strain. It includes the material's elastic constants and coupling coefficients as:
\be
\mathbf{C}^{(p)} =
\begin{bmatrix}
\mu_{11}^{(p)} & \mu_{12}^{(p)} & \mu_{13}^{(p)} & 0 & 0 & 0 & 0 & 0 & -e_{31}^{(p)} \\
\mu_{12}^{(p)} & \mu_{11}^{(p)} & \mu_{13}^{(p)} & 0 & 0 & 0 & 0 & 0 & -e_{31}^{(p)} \\
\mu_{13}^{(p)} & \mu_{13}^{(p)} & \mu_{33}^{(p)} & 0 & 0 & 0 & 0 & 0 & -e_{33}^{(p)} \\
0 & 0 & 0 & \mu_{44}^{(p)} & 0 & 0 & 0 & -e_{15}^{(p)} & 0 \\
0 & 0 & 0 & 0 & \mu_{44}^{(p)} & 0 & -e_{15}^{(p)} & 0 & 0 \\
0 & 0 & 0 & 0 & 0 & \mu_{44}^{(p)} & 0 & 0 & 0 \\
0 & 0 & 0 & 0 & e_{15}^{(p)} & 0 & \kappa_{11}^{(p)} & 0 & 0 \\
0 & 0 & 0 & e_{15}^{(p)} & 0 & 0 & 0 & \kappa_{11}^{(p)} & 0 \\
e_{31}^{(p)} & e_{31}^{(p)} & e_{33}^{(p)} & 0 & 0 & 0 & 0 & 0 & \kappa_{33}^{(p)}
\end{bmatrix},
\label{C(p)}
\ee
%%%-------------------------------------
the expression of strain component$ (S^{(p)}_{ij})$ and external field $(E^{(p)}_{i})$ is given as:
\be
 S^{(p)}_{ij} = \frac{1}{2} \left( u^{(p)}_{i,j}+u^{(p)}_{j,i} \right) , \:\:\: E^{(p)}_{i}= -\frac{\partial \phi^{(p)}}{\partial x_i} .
 \label{straon_relation}
\ee
 In Eq. (\ref{straon_relation}), $\phi^{(p)}$ denotes the electric potential; and the material parameters $(\mu^{(p)}_{44}$, $e^{(p)}_{15}$, $\kappa^{(p)}_{11})$ are defined as:
\be
\mu^{(p)}_{44}=(\mu^{(p)}_{44})_e+ i\omega (\mu^{(p)}_{44})_v, \:\:\: e^{(p)}_{15}=(e^{(p)}_{15})_e+i \omega (e^{(p)}_{15})_v, \:\:\: \kappa^{(p)}_{11}=(\kappa^{(p)}_{11})_e+i \omega (\kappa^{(p)}_{11})_v,
\label{eq2}
\ee
where, $(\mu^{(p)}_{44})_e$ and $(\mu^{(p)}_{44})_v$ are elastic and viscoelastic constants; $(e^{(p)}_{15})_e$ and $(e^{(p)}_{15})_v$ are piezoelectric constant and piezoelectric loss moduli; $(\kappa^{(p)}_{11})_e$ and $(\kappa^{(p)}_{11})_v$ are dielectric constants and dielectric loss moduli. Consequently, by using Eqs. (\ref{displacement_component}), (\ref{constitutive relation})–(\ref{straon_relation}), the constitutive relations for a transversely isotropic piezoelectric-viscoelastic material half-space under anti-plane deformation can be represented as follows:
\be
\left.
\ba{lll}
\begin{bmatrix}
T^{(p)}_{xz} \\
T^{(p)}_{yz}
\end{bmatrix}=
\begin{bmatrix}
\mu^{(p)}_{44} \frac{\partial}{\partial x} & e^{(p)}_{15} \frac{\partial}{\partial x} \\
 \mu^{(p)}_{44}\frac{\partial}{\partial y} & e^{(p)}_{15} \frac{\partial}{\partial y}
\end{bmatrix}
\begin{bmatrix}
w^{(p)} \\
\phi^{(p)}
\end{bmatrix},
 \:\:\: \begin{bmatrix}
D^{(p)}_x \\
D^{(p)}_{y}
\end{bmatrix}=
\begin{bmatrix}
e^{(p)}_{15} \frac{\partial}{\partial x} & -\kappa^{(p)}_{11} \frac{\partial}{\partial x} \\
e^{(p)}_{15} \frac{\partial}{\partial y} & -\kappa^{(p)}_{11} \frac{\partial}{\partial y}
\end{bmatrix}
\begin{bmatrix}
w^{(p)} \\
\phi^{(p)}
\end{bmatrix}.
\ea
\right.
\label{eq1}
\ee
The bimaterial composite structure is considered to rotates about the \( y \)-axis with a uniform angular velocity $\mathbf{\Re^{(p)}}$, given by
$ \mathbf{\Re^{(p)}} = (0, \: \Re^{(p)}, \: 0).$
It is also assumed that there is no initial electric displacement vector, i.e.,
$D^{0(p)}_j = 0$.
 Furthermore, the constant pre-stress component \(\Im^{(p)}_{kj}\) is considered to be present only in the \( y \)-direction.\\
To differentiate between the stress components, mechanical and electrical displacement components, and material constants for the two half-spaces, subscripts \( \text{u} \) and \( \text{l} \) are used, where \( \text{u} \) denotes the FGPVM half-space and \( \text{l} \) denotes the PVM half-space.

\subsubsection{Dynamics for FGPVM (upper half-space)}
% Employing Eq. \ref{Equation for piezoelectric layer}, the governing equation of motion as well as the electric field equation for initially stressed FGPVM half-space, having no distributed body force, subjected to rotation along the $y$-axis is given by:
The governing equation of motion and the electric field equation for initially stressed upper half-space (FGPVM) are derived from Eq. (\ref{Equation for piezoelectric layer})  as follows:
\be
\mathbf{M} \cdot \mathbf{X} = \mathbf{B},
\label{eq3}
\ee
where,
matrix \(\mathbf{M}\) organizes the coefficients and terms associated with \(T^{(u)}_{xz,x}\), \(T^{(u)}_{yz,y}\), \(w^{(u)}_{,yy}\), \(D^{(u)}_{x,x}\), and \(D^{(u)}_{y,y}\), as:

\[
\mathbf{M} = \begin{bmatrix}
1 & 1 & \Im^{(u)} & 0 & 0 \\
0 & 0 & 0 & 1 & 1
\end{bmatrix},
\]
vector \(\mathbf{X}\) contains the field variables while vector \(\mathbf{B}\) represents the non-homogeneous part of the equations given by:
\[
\mathbf{X} = 
\begin{bmatrix}
T^{(u)}_{xz,x} & T^{(u)}_{yz,y} & w^{(u)}_{,yy} & D^{(u)}_{x,x} & D^{(u)}_{y,y}
\end{bmatrix}^{\text{T}}
,  \:\: 
\mathbf{B} = \begin{bmatrix}
\rho^{(u)} \left( w^{(u)}_{,tt} - (\Re^{(u)})^2 w^{(u)} \right) \\
0
\end{bmatrix},
\]
where, $\Im^{(u)}(=\Im^{(u)}_{yy})$ is initial stress corresponding to upper half-space.  The notation \((\cdot)_{,i}\) denotes the differentiation of \((\cdot)\) with respect to the variable \(x_i\). The material parameters for the upper half-space are modeled with an exponential depth profile having $\xi_1$ as the heterogeneity parameter, considered as: 
\be
\begin{aligned}
\left[ \: (\mu^{(u)}_{44})_e(x),\:  (\mu^{(u)}_{44})_v(x), \: (e^{(u)}_{15})_e(x),\: (e^{(u)}_{15})_v(x), \: (\kappa^{(u)}_{11})_e(x),\: (\kappa^{(u)}_{11})_v(x),\: \rho^{(u)}(x),\: \Im^{(u)}(x) \: \right] \\
=\left[ \:(\mu^{(u)}_{44})_e^{'},\:  (\mu^{(u)}_{44})_v^{'}, \: (e^{(u)}_{15})_e^{'},\: (e^{(u)}_{15})_v^{'}, \: (\kappa^{(u)}_{11})_e^{'},\: (\kappa^{(u)}_{11})_v^{'},\: {{\rho^{(u)}}^{'}}, \: {{\Im^{(u)}}^{'}} \: \right] \:e^{\xi_1 x},
\end{aligned}
\label{eq4}
\ee
% where $\xi_1$ is the exponential gradient parameter.
% Using Eqs. (\ref{eq2}), (\ref{eq1}) and (\ref{eq4}) in Eq. (\ref{eq3}), which reduces to the following form:
Using Eqs. (\ref{eq2}), (\ref{eq1}), and (\ref{eq4}), Eq. (\ref{eq3}) is transformed into the following form:
\begin{equation}
\begin{aligned}
\begin{bmatrix}
\begin{array}{l}
\overline{\mu^{(u)}_{44}} \frac{\partial^2}{\partial x^2} + \left[ \overline{\mu^{(u)}_{44}} + {\Im^{(u)}}' \right] \frac{\partial^2}{\partial y^2} + \xi_1 \overline{\mu^{(u)}_{44}} \frac{\partial}{\partial x} - {\rho^{(u)}}' \frac{\partial^2}{\partial t^2} + {\rho^{(u)}}' (\Re^{(u)})^2 \\
\overline{e^{(u)}_{15}} \left( \frac{\partial^2}{\partial x^2} + \frac{\partial^2}{\partial y^2} + \xi_1 \frac{\partial}{\partial x} \right)
\end{array}
&
\begin{array}{l}
\overline{e^{(u)}_{15}} \left( \frac{\partial^2}{\partial x^2} + \frac{\partial^2}{\partial y^2} + \xi_1 \frac{\partial}{\partial x} \right) \\
-\overline{\kappa^{(u)}_{11}} \left( \frac{\partial^2}{\partial x^2} + \frac{\partial^2}{\partial y^2} + \xi_1 \frac{\partial}{\partial x} \right)
\end{array}
\end{bmatrix}
\begin{bmatrix}
w^{(u)} \\
\phi^{(u)}
\end{bmatrix}
=\begin{bmatrix}
0 \\
0
\end{bmatrix}
\end{aligned}
\label{matrix_form_eq5and6}
\end{equation}
where, 
\begin{equation*}
\overline{\mu^{(u)}_{44}}=(\mu^{(u)}_{44})_e^{'}+i \: \omega \: (\mu^{(u)}_{44})_v^{'}, \:\:\:
\overline{e^{(u)}_{15}}\:=\: (e^{(u)}_{15})_e^{'}+i \: \omega\: (e^{(u)}_{15})_v^{'}, \:\:\: 
\overline{\kappa^{(u)}_{11}}\:=\:(\kappa^{(u)}_{11})_e^{'}+i \; \omega\: (\kappa^{(u)}_{11})_v^{'}.
\label{eq7}
\end{equation*}
Equation (\ref{matrix_form_eq5and6}) represents the governing equation for upper half-space in terms of displacement and electric potential. In the next subsection, the dynamics for the lower half-space will be discussed.
\subsubsection{Dynamics for PVM (lower half-space)}
The governing equation of motion and electric field equations for lower half-space (PVM) in the absence of pre-stress is derived from Eq. (\ref{Equation for piezoelectric layer}) as follows:
\be
\mathbf{M^{'}} \cdot \mathbf{X{'}} = \mathbf{B{'}},
\label{eq8}
\ee
where, matrix $\mathbf{M^{'}}$ organizes the coefficient and terms associated with $T^{(l)}_{xz,x}, \:
T^{(l)}_{yz,y},\:
D^{(l)}_{x,x},$ and
$D^{(l)}_{y,y}$ as:
\[
\mathbf{M^{'}} =
\begin{bmatrix}
1 & 1 & 0 & 0 \\
0 & 0 & 1 & 1
\end{bmatrix},
\] 
vector $\mathbf{X^{'}}$ contains the field variables while vector $\mathbf{B^{'}}$ represents the non-homogeneous part of the equations given by:
\[
\mathbf{X'} = 
\begin{bmatrix}
T^{(l)}_{xz,x} & T^{(l)}_{yz,y} & D^{(l)}_{x,x} & D^{(l)}_{y,y}
\end{bmatrix}^{\text{T}}
, \:\:\:
\mathbf{B^{'}} =
\begin{bmatrix}
\rho^{(l)} \left[ w^{(l)}_{,tt} - (\Re^{(l)})^2 w^{(l)} \right] \\
0
\end{bmatrix},
\]
where, $\rho^{(l)}$ and $\Re^{(l)}$ represent density and rotation parameters for the  considered layer.
Equation (\ref{eq8}) is reduced into the following form with the help of  Eqs. (\ref{eq2})  and (\ref{eq1}) as:
\begin{equation}
\begin{aligned}
\begin{bmatrix}
\begin{array}{l}
\mu^{(l)}_{44} \left( \frac{\partial^2}{\partial x^2} + \frac{\partial^2}{\partial y^2} \right) \\
e^{(l)}_{15} \left( \frac{\partial^2}{\partial x^2} + \frac{\partial^2}{\partial y^2} \right)
\end{array}
&
\begin{array}{l}
e^{(l)}_{15} \left( \frac{\partial^2}{\partial x^2} + \frac{\partial^2}{\partial y^2} \right) \\
-\kappa^{(l)}_{11} \left( \frac{\partial^2}{\partial x^2} + \frac{\partial^2}{\partial y^2} \right)
\end{array}
\end{bmatrix}
\begin{bmatrix}
w^{(l)} \\
\phi^{(l)}
\end{bmatrix}
=
\begin{bmatrix}
\rho^{(l)} \left( \frac{\partial^2}{\partial t^2} - (\Re^{(l)})^2 \right) w^{(l)} \\
0
\end{bmatrix}.
\label{matrix_form_eq9_10}
\end{aligned}
\end{equation}
Equation (\ref{matrix_form_eq9_10}) represents the governing equation for lower half-space in terms of displacement and electric potential. The solutions for the Eqs. (\ref{matrix_form_eq5and6}, \ref{matrix_form_eq9_10}) will be obtained in the following subsection.
\subsection{Solution of the field variables}
\label{Solution of the field variables}
\subsubsection{FGPVM (upper half-space) solution}
In Eq.~(\ref{matrix_form_eq5and6}), $w^{(u)}$ and $\phi^{(u)}$ are initially coupled. However, after substituting Eq.~(\ref{eq11}), this coupling is transformed, resulting in a system where $w^{(u)}$ and $\phi^{(u)}$ are uncoupled.

\be
\varphi^{(u)}=\phi^{(u)}-w^{(u)} \frac{\overline{e^{(u)}_{15}}}{ \overline{\kappa^{(u)}_{11}}}.
\label{eq11}
\ee
Substituting Eq. (\ref{eq11}) into Eq. (\ref{matrix_form_eq5and6}), the resulting form is obtained as:
\be
\mathbf{N} \cdot \mathbf{Y} = \mathbf{C},
\label{matrix_eq15}
\ee
where, matrix \(\mathbf{N}\) captures the coefficients and operators applied to \(w^{(u)}\) and \(\varphi^{(u)}\) given by:
 \be
\mathbf{N} =  \begin{bmatrix} \left[ \overline{\mu^{(u)}_{44}} + \frac{(\overline{e^{(u)}_{15}})^2}{\overline{\kappa^{(u)}_{11}}} \right] \frac{\partial^2}{\partial x^2} + \left[ \overline{\mu^{(u)}_{44}} + {\Im^{(u)}}' + \frac{(\overline{e^{(u)}_{15}})^2}{\overline{\kappa^{(u)}_{11}}} \right] \frac{\partial^2}{\partial y^2} + \xi_1 \left[ \overline{\mu^{(u)}_{44}} + \frac{(\overline{e^{(u)}_{15}})^2}{\overline{\kappa^{(u)}_{11}}} \right] \frac{\partial}{\partial x} & 0 \\
0 & \frac{\partial^2}{\partial x^2} +\frac{\partial^2 }{\partial y^2}+\xi_1 \: \frac{\partial }{\partial x} \end{bmatrix},
\ee
 vector \(\mathbf{Y}\) contains the functions \(w^{(u)}\) and \(\varphi^{(u)}\), whereas vector \(\mathbf{C}\) represents the non-homogeneous part of the equations given by:
\be
\mathbf{Y} = \begin{bmatrix}
w^{(u)} \\
\varphi^{(u)}
\end{bmatrix}, \:\:\:
\mathbf{C} = \begin{bmatrix}
\rho^{(u)}{}^{'} \left( \frac{\partial^2 w^{(u)}}{\partial t^2} - (\Re^{(u)})^2 w^{(u)} \right) \\
0
\end{bmatrix}.
\ee
The Love wave travels with a constant velocity \( c \) in the \( y \)-direction, and the coordinate system \((x,\: y,\: z)\) is replaced by \((x_1, \:x_2, \:x_3)\), which is oriented in the direction of wave propagation. Consequently, by Galilean transformation law:
\be
 x_1=x, \:\:\:x_2=y-ct \:\:\: x_3=z.
 \label{eq14}
 \ee
 By applying the transformation \(x_2 = y - ct\), it follows that \({\partial}/{\partial y} = {\partial}/{\partial x_2}\) and \({\partial}/{\partial t} = c \: {\partial}/{\partial x_2}\). Consequently, the mechanical displacement and electric potential become functions of \(x_1\), \(x_2\), and \(t\). Thus, Eq. (\ref{matrix_eq15}) can be reformulated in terms of these moving coordinates as follows:
\begin{equation}
\begin{bmatrix}
\beta^{(u)}_f \frac{\partial^2}{\partial x_1^2} + \left[\beta^{(u)}_f + \Im^{(u)'} - \rho^{(u)'} c^2 \right] \frac{\partial^2}{\partial x_2^2} + \xi_1 \beta^{(u)}_f \frac{\partial}{\partial x_1} + \rho^{(u)'} (\Re^{(u)})^2 & 0 \\
0 & \frac{\partial^2}{\partial x_1^2} + \frac{\partial^2}{\partial x_2^2} + \xi_1 \frac{\partial}{\partial x_1}
\end{bmatrix}
\begin{bmatrix}
w^{(u)} \\
\varphi^{(u)}
\end{bmatrix}
=
\begin{bmatrix}
0 \\
0
\end{bmatrix}
\label{matrix_1516}
\end{equation}
where,
\begin{equation*}
\beta^{(u)}_f=\left[ \: \overline{\mu^{(u)}_{44}} + \: \frac{(\overline{e^{(u)}_{15}})^2}{\overline{\kappa^{(u)}_{11}}}\: \right].
\label{eq17}
\end{equation*}
The integral solutions corresponding to expression (\ref{matrix_1516}) are assumed to take the following form:
%%%%% matrix form

\be
\begin{bmatrix}
w^{(u)}(x_1, x_2) \\
\varphi^{(u)}(x_1, x_2)
\end{bmatrix}
= \int_0^\infty \mathbf{P}(\eta, x_1) \cdot \mathbf{Q}(\eta, x_2) \, d\eta,
\label{eq18}
\ee
where, 
\begin{equation*}
\mathbf{P}(\eta, x_1) =
\begin{bmatrix}
K(\eta, x_1) & L(\eta, x_1) \\
M(\eta, x_1) & N(\eta, x_1)
\end{bmatrix}, \:\:\:
\mathbf{Q}(\eta, x_2) =
\begin{bmatrix}
\text{Re}(e^{i \eta x_2}) \\
\text{Im}(e^{i \eta x_2})
\end{bmatrix},
\end{equation*}
where, $K(\eta,\: x_1 )$, $L (\eta , \:x_1 )$, $M(\eta, \:x_1 )$ and $N(\eta, \:x_1 )$ are unknown functions to be determined.
By substituting the values from Eqs. (\ref{eq18}) into Eq. (\ref{matrix_1516}), and assuming that both shear stress and electric displacement approach zero at the boundary (\(x_1 \to \infty\)) while remaining bounded, i.e.,
\be
T^{(u)}_{13}(\infty, \: x_2)=0, \:\:\: D^{(u)}_{1}(\infty, \: x_2)=0.
\label{eq19}
\ee 
The final expressions for mechanical displacement and electric potential are as follows:
\be
\begin{bmatrix}
\displaystyle w^{(u)}(x_1, x_2) \\
\displaystyle \varphi^{(u)}(x_1, x_2)
\end{bmatrix}
= \int_0^\infty \mathbf{U}(\eta,x_1) \cdot \mathbf{V}(\eta, x_2) \, d\eta
+ \mathbf{W} \cdot w^{(u)}(x_1, x_2),
\label{eq21}
\ee
where,
 \begin{equation*}
\mathbf{U}(\eta, x_1) =
\begin{bmatrix}
K_1(\eta) \: e^{-\lambda_1 x_1} & L_1(\eta) \: e^{-\lambda_1 x_1}\\
M_1(\eta) \:  e^{-\chi_1 x_1} & N_1(\eta) \: e^{-\chi_1 x_1}
\end{bmatrix}
, \:\:\: \mathbf{V}(\eta, x_2) =
\begin{bmatrix}
\text{Re}(e^{i \eta x_2}) \\
\text{Im}(e^{i \eta x_2})
\end{bmatrix}, \:\:\: \mathbf{W} =
\begin{bmatrix}
0 \\
\frac{\overline{e^{(u)}_{15}}}{\overline{\kappa^{(u)}_{11}}}
\end{bmatrix},
 \end{equation*}
\begin{align*}
\lambda_1 = \frac{\xi_1 + \sqrt{\xi_1^2 + 4 \left[ \left( 1 + \frac{{\Im^{(u)}}^{'}}{\beta^{(u)}_f} - \frac{c^2}{{(\beta^{(u)}})^2} \right) \eta^2 - \frac{(\Re^{(u)})^2}{{(\beta^{(u)})^2}} \right]}}{2},\:\:
\chi_1=\frac{\xi_1+\sqrt{\xi^2_1+4 \eta^2}}{2},\:\:
{\beta^{(u)}}&=\sqrt{\frac{\beta^{(u)}_f}{\overline{\rho^{(u)}}}}.
\end{align*}
Here $\beta^{(u)}$ is Love wave velocity associated with FGPVM half-space and K$_1$, L$_1$, M$_1$ and  N$_1$ are unknown functions.
Using Eqs. (\ref{eq21}), the stress and electric components for FGPVM half-space, as defined in Eq. (\ref{eq1}), are given by:
\be
T_{13}^{(u)} = \left( -\beta_f^{(u)} \mathbf{S} - \overline{e^{(u)}_{15}} \mathbf{T} \right) e^{\xi_1 x_1}, \:\:\: D^{(u)}_1= \left( \: \overline{\kappa^{(u)}_{11}} \mathbf{T} \: \right) e^{\xi_1 x_1},
\label{eq25}
\ee
\be
T_{23}^{(u)} = \left( \beta_f^{(u)} \mathbf{E} + \overline{e^{(u)}_{15}} \mathbf{F} \right) e^{\xi_1 x_1}, \:\:\: D^{(u)}_2= -\left( \: \overline{\kappa^{(u)}_{11}} \mathbf{F} \: \right) e^{\xi_1 x_1},
\label{eq26}
\ee
where,
 \begin{equation*}
\mathbf{S} = \int_0^\infty \lambda_1 \left( \mathbf{f}(\eta)^T \mathbf{h}(\eta) \right) e^{-\lambda_1 x_1} \, d\eta, \quad
\mathbf{T} = \int_0^\infty \chi_1 \left( \mathbf{g}(\eta)^T \mathbf{h}(\eta) \right) e^{-\chi_1 x_1} \, d\eta,
\end{equation*}
 \begin{equation*}
\mathbf{E} = \int_0^\infty \eta \left( \mathbf{l}(\eta)^T \mathbf{n}(\eta) \right) e^{-\lambda_1 x_1} \, d\eta, \quad
\mathbf{F} = \int_0^\infty \eta \left( \mathbf{m}(\eta)^T \mathbf{n}(\eta) \right) e^{-\chi_1 x_1} \, d\eta,
\end{equation*}
 \begin{equation*}
\mathbf{f}(\eta) = \begin{bmatrix}
K_1(\eta) \\
L_1(\eta)
\end{bmatrix}, \quad
\mathbf{g}(\eta) = \begin{bmatrix}
M_1(\eta) \\
N_1(\eta)
\end{bmatrix}, \quad
\mathbf{h}(\eta) = \begin{bmatrix}
\text{Re}(e^{i \eta x_2}) \\
\text{Im}(e^{i \eta x_2})
\end{bmatrix},
 \end{equation*}
 \begin{equation*}
\mathbf{l}(\eta) = \begin{bmatrix}
-K_1(\eta) \\
L_1(\eta)
\end{bmatrix}, \quad
\mathbf{m}(\eta) = \begin{bmatrix}
-M_1(\eta) \\
N_1(\eta)
\end{bmatrix}, \quad
\mathbf{n}(\eta) = \begin{bmatrix}
\text{Im}(e^{i \eta x_2}) \\
\text{Re}(e^{i \eta x_2})
\end{bmatrix}.
\end{equation*} 

\subsubsection{PVM (lower half-space) solution}
 In Eq.~(\ref{matrix_form_eq9_10}), \( w^{(l)} \) and \( \phi^{(l)} \) are initially coupled, meaning they are interdependent in the equation. However, after applying the substitution from Eq.~(\ref{eq29}), this relationship changes, and \( w^{(l)} \) and \( \phi^{(l)} \) become decoupled, resulting in two separate equations where they no longer directly influence each other.

\be
\varphi^{(l)}=\phi^{(l)}-w^{(l)} \frac{e^{(l)}_{15}}{\kappa^{(l)}_{11}}.
\label{eq29}
\ee
Substituting Eq. (\ref{eq29}) in Eq. (\ref{matrix_form_eq9_10}) and solving, the following result is obtained:
\be
\mathbf{N^{'}} \cdot \mathbf{Y^{'}} = \mathbf{C^{'}},
\label{eq30}
\ee
where, matrix $\mathbf{N^{'}}$ represents  the coefficient and operator applied to $w^{(l)}$ and $\phi^{(l)}$ given by:
\[
\mathbf{N^{'}} =
\begin{bmatrix}
\beta^{(l)}_f \left( \frac{\partial^2}{\partial x^2} + \frac{\partial^2}{\partial y^2} \right) & 0 \\
0 & \left( \frac{\partial^2}{\partial x^2} + \frac{\partial^2}{\partial y^2} \right)
\end{bmatrix},
\]
vector $\mathbf{Y^{'}}$ contains the function  $w^{(l)}$ and $\phi^{(l)}$, and vector $\mathbf{C^{'}}$ represents the non-homogenous part of the equation given by:
\[
\mathbf{Y^{'}} =
\begin{bmatrix}
 w^{(l)} \\
 \varphi^{(l)}
\end{bmatrix}, \:\:\:
\mathbf{C^{'}} =
\begin{bmatrix}
{\rho^{(l)}} \left( \frac{\partial^2 w^{(l)}}{\partial t^2} - {(\Re^{(l)})^2} w^{(l)} \right) \\
0
\end{bmatrix},
\]
where,
\begin{equation*}
\displaystyle \beta^{(l)}_f=\left[ \: \mu^{(l)}_{44} + \: \frac{({e^{(l)}_{15}})^2}{{\kappa^{(l)}_{11}}}\: \right].
\label{eq31}
\end{equation*}
By applying the Galilean transformation defined earlier in Eq. (\ref{eq14}), Eqs. (\ref{eq30}) are transformed to:
\be
\begin{bmatrix}
\beta^{(l)}_f \frac{\partial^2}{\partial x_1^2} + (\beta^{(l)}_f - \rho^{(l)} c^2) \frac{\partial^2}{\partial x_2^2} + \rho^{(l)} (\Re^{(l)})^2 & 0 \\
0 & \frac{\partial^2}{\partial x_1^2} + \frac{\partial^2}{\partial x_2^2}
\end{bmatrix}
\begin{bmatrix}
w^{(l)} \\
\varphi^{(l)}
\end{bmatrix}
=
\begin{bmatrix}
0 \\
0
\end{bmatrix}.
\label{eq32}
\ee
Assuming the solution of Eq.~ (\ref{eq32}) in integral form as:
\be
\begin{bmatrix}
w^{(l)}(x_1, x_2) \\
\varphi^{(l)}(x_1, x_2)
\end{bmatrix}
= \int_0^\infty \mathbf{P^{'}}(\eta, x_1) \cdot \mathbf{Q^{'}}(\eta, x_2) \, d\eta,
\label{eq33}
\ee
where, 
 \begin{equation*}
\mathbf{P^{'}}(\eta, x_1) =
\begin{bmatrix}
K^{'}(\eta, x_1) & L^{'}(\eta, x_1) \\
M^{'}(\eta, x_1) & N^{'}(\eta, x_1)
\end{bmatrix}, \:\:\:
\mathbf{Q^{'}}(\eta, x_2) =
\begin{bmatrix}
\text{Re}(e^{i \eta x_2}) \\
\text{Im}(e^{i \eta x_2})
\end{bmatrix},
 \end{equation*}
where, $K^{'}(\eta,\: x_1 )$, $L^{'} (\eta , \:x_1 )$, $M^{'}(\eta, \:x_1 )$ and $N^{'}(\eta, \:x_1 )$ are functions yet to be determined. Substitute the value of Eq. (\ref{eq33}) in Eq. (\ref{eq32})  and consider the assumption that the shear stress and electric displacement approach zero as $x_1 \to -\infty$, namely:
\be
T^{(l)}_{13}(-\infty, \: x_2)=0, \:\:\: D^{(l)}_{1}(-\infty, \: x_2)=0.
\label{eq34}
\ee 
The final expressions for mechanical displacement and electric potential are as follows:
\be
\begin{bmatrix}
\displaystyle w^{(l)}(x_1, x_2) \\
\displaystyle \varphi^{(l)}(x_1, x_2)
\end{bmatrix}
= \int_0^\infty \mathbf{U^{'}}(\eta,x_1) \cdot \mathbf{V^{'}}(\eta, x_2) \, d\eta
+ \mathbf{W^{'}} \cdot w^{(l)}(x_1, x_2),
\label{eq35}
\ee
where,
\begin{equation*}
\mathbf{U'}(\eta, x_1) =
\begin{bmatrix}
K_2(\eta) \: e^{\gamma_2 x_1} & L_2(\eta) \: e^{\gamma_2 x_1} \\
M_2(\eta) \: e^{\eta x_1} & N_2(\eta) \: e^{\eta x_1}
\end{bmatrix}, \:\:\: \mathbf{V^{'}}(\eta, x_2) =
\begin{bmatrix}
\text{Re}(e^{i \eta x_2}) \\
\text{Im}(e^{i \eta x_2})
\end{bmatrix}, \:\:\: \mathbf{W^{'}} =
\begin{bmatrix}
0 \\
\frac{{e^{(l)}_{15}}}{{\kappa^{(l)}_{11}}}
\end{bmatrix},
\end{equation*}

 \begin{equation*}
\left.
\ba{lll}
\displaystyle \gamma_2&=\sqrt{\eta^2 \left( 1-\frac{c^2}{{(\beta^{(l)}})^2} \right)-\frac{{\Re^{(l)}}^2}{(\beta^{(l)})^2}},\:\:\:
\displaystyle {\beta^{(l)}}&=\sqrt{\frac{\beta^{(l)}_f}{{\rho^{(l)}}}}.
\ea
\right.
\label{eq37}
\end{equation*}
Here, \(\beta^{(l)}\) represents the Love wave velocity associated with the PVM half-space, and \( K_2(\eta) \), \( L_2(\eta) \), \( M_2(\eta) \), and \( N_2(\eta) \) are functions that need to be determined.
From the expression provided in Eq.~(\ref{eq35}), the stress and electric displacement components for the PVM half-space, as defined in Eq.~(\ref{eq1}), are computed as follows:

\be
T_{13}^{(l)} = \beta_f^{(l)} \mathbf{S^{'}} + {e^{(l)}_{15}} \mathbf{T^{'}},\:\:\: D^{(l)}_1=  \: -{\kappa^{(l)}_{11}} \mathbf{T^{'}}, \:  
\label{eq38}
\ee
 \be
T_{23}^{(l)} =  \beta_f^{(l)} \mathbf{E^{'}} + {e^{(l)}_{15}} \mathbf{F^{'}}, \:\:\: D^{(l)}_2= -\: {\kappa^{(l)}_{11}} \mathbf{F^{'}}, \:  
 \label{stressT13(l)}
\ee
where,
\begin{equation*}
\mathbf{S^{'}} = \int_0^\infty \gamma_2 \left( \mathbf{f^{'}}(\eta)^T \mathbf{h^{'}}(\eta) \right) e^{\gamma_2 x_1} \, d\eta, \quad
\mathbf{T^{'}} = \int_0^\infty \eta \left( \mathbf{g^{'}}(\eta)^T \mathbf{h^{'}}(\eta) \right) e^{\eta x_1} \, d\eta,
\end{equation*}
\begin{equation*}
 \mathbf{E^{'}} = \int_0^\infty \eta \left( \mathbf{l^{'}}(\eta)^T \mathbf{n^{'}}(\eta) \right) e^{ \gamma_2 x_1} \, d\eta, \quad
 \mathbf{F^{'}} = \int_0^\infty \eta \left( \mathbf{m^{'}}(\eta)^T \mathbf{n^{'}}(\eta) \right) e^{\eta x_1} \, d\eta,
 \end{equation*}
 
 \begin{equation*}
 \mathbf{f^{'}}(\eta) = \begin{bmatrix}
 K_2(\eta) \\
 L_2(\eta)
 \end{bmatrix}, \quad
 \mathbf{g^{'}}(\eta) = \begin{bmatrix}
 M_2(\eta) \\
 N_2(\eta)
 \end{bmatrix}, \quad
 \mathbf{h^{'}}(\eta) = \begin{bmatrix}
 \text{Re}(e^{i \eta x_2}) \\
 \text{Im}(e^{i \eta x_2})
 \end{bmatrix},
\end{equation*}
  \begin{equation*}
  \mathbf{l^{'}}(\eta) = \begin{bmatrix}
  -K_2(\eta) \\
  L_2(\eta)
  \end{bmatrix}, \quad
  \mathbf{m{'}}(\eta) = \begin{bmatrix}
  -M_2(\eta) \\
  N_2(\eta)
  \end{bmatrix}, \quad
  \mathbf{n^{'}}(\eta) = \begin{bmatrix}
  \text{Im}(e^{i \eta x_2}) \\
  \text{Re}(e^{i \eta x_2})
  \end{bmatrix}. 
\end{equation*} 

\subsection{Boundary conditions}
\label{section bc}
In the analysis of the crack, several boundary conditions are imposed to ensure the system's accurate physical behavior. These conditions address both the mechanical and electrostatic properties of the crack surfaces. \\
The elastic and electrostatic boundary conditions at the crack surfaces are:
\be
T^{(u)}_{13}(0^{+},\:x_2)=T^{(l)}_{13}(0^{-},\:x_2)=-T_0, \:\:\: D^{(u)}_1(0^{+},\:x_2)=D^{(l)}_1(0^{-},\:x_2) =D_0,\:\:\:\:  |x_2| \le d.
\label{bc1}
\ee
The crack is assumed to be under anti-plane deformation, with frictionless upper and lower surfaces. Therefore, the crack can be treated as electrically permeable, leading to the following conditions:
\be
D^{(u)}_1(0^{+},\:x_2)=D^{(l)}_1(0^{-},\:x_2),\:\:\:  \phi^{(u)}(0^{+},\:x_2)=\phi^{(l)}(0^{-},\:x_2), \:\:\:\: |x| < \infty.
\label{bc2}
\ee
Additionally, the elastic and electric fields must fulfill the continuity conditions along the crack interface as described:
\be
 w^{(u)}(0^{+},\:x_2)= w^{(l)}(0^{-},\:x_2), \:\:\: \phi^{(u)}(0^{+},\:x_2)=\phi^{(l)}(0^{-},\:x_2), \:\:\:\: |x_2| > d,
 \label{bc3}
\ee
\be
T^{(u)}_{13}(0^{+},\:x_2)=T^{(l)}_{13}(0^{-},\:x_2), \:\:\: D^{(u)}_1(0^{+},\:x_2)=D^{(l)}_1(0^{-},\:x_2),\:\:\:\:\:\:\: |x_2| > d.
\label{bc4}
\ee

\subsection{Formulation of SIF and EDIF}
\label{Formulation of Stress Intensity and Electric Displacement Intensity Factors}
 Using the boundary conditions (\ref{bc1})-(\ref{bc4}), the following results are obtained:
\be
\begin{aligned}
\displaystyle  \int_0^\infty  \lambda_1 \:\left[ K_1(\eta) \: \text{Re} (e^{(i \eta x_2)}) +L_1 (\eta ) \: \text{Im} (e^{(i \eta x_2)})\: \right] \: d\eta 
 \displaystyle = \frac{1}{\beta^{(u)}_f} \:\left[ T_0-\frac{\overline{e^{(u)}_{15}}}{\overline{\kappa^{(u)}_{11}}}  D_0\right], \:\:\: |x_2| \le d,
\end{aligned}
\label{eq43}
\ee
\be
\begin{aligned}
\displaystyle  \int_0^\infty  \left[ \:( K_1(\eta )-K_2(\eta)) \: \text{Re} (e^{(i \eta x_2)}) +(L_1 (\eta) -L_2 (\eta )) \: \text{Im} (e^{(i \eta x_2)})\: \right] \: d\eta 
 \displaystyle = 0, \:\:\: |x_2| > d,
\end{aligned}
\label{eq44}
\ee
\be
\begin{aligned}
\displaystyle  \int_0^\infty  \chi_1 \:\left[ M_1(\eta) \: \text{Re} (e^{(i \eta x_2)}) +N_1 (\eta ) \: \text{Im} (e^{(i \eta x_2)})\: \right] \: d\eta 
 \displaystyle = \frac{D_0}{\overline{\kappa^{(u)}_{11}}}, \:\:\: |x_2| \le d,
\end{aligned}
\label{eq45}
\ee
\be
\int_0^\infty \left[ \left(M_1(\eta)+\frac{\overline{e^{(u)}_{15}}}{\overline{\kappa^{(u)}_{11}}} K_1(\eta) \right) -\left(M_2(\eta)+\frac{e^{(l)}_{15}}{\kappa^{(l)}_{11}} K_2(\eta) \right) \right]\text{Re} (e^{(i \eta x_2)})  d\eta =0,  \:\:\: |x_2| > d,
\label{eq46}
\ee
\be
\int_0^\infty \left[ \left(N_1(\eta)+\frac{\overline{e^{(u)}_{15}}}{\overline{\kappa^{(u)}_{11}}} L_1(\eta) \right) -\left(N_2(\eta)+\frac{e^{(l)}_{15}}{\kappa^{(l)}_{11}} L_2(\eta) \right) \right]\text{Im} (e^{(i \eta x_2)})  d\eta =0,  \:\:\: |x_2| > d.
\label{eq47}
\ee
 By examining the above equation and differentiating between the even and odd functions of \(x_2\), which correspond to the symmetric and anti-symmetric components relative to \(x_2\), it can be determined that:

 \be
\begin{aligned}
\displaystyle  \int_0^\infty  \lambda_1 \: K_1(\eta) \: \text{Re} (e^{(i \eta x_2)})  \: d\eta = \frac{1}{\beta^{(u)}_f} \:\left[ T_0-\frac{\overline{e^{(u)}_{15}}}{\overline{\kappa^{(u)}_{11}}}  D_0\right], \:\:\: 0<x_2< d,
\end{aligned}
\label{eq48}
\ee
\be
\begin{aligned}
\displaystyle  \int_0^\infty   \:[ K_1(\eta )-K_2(\eta) ]\: \text{Re} (e^{(i \eta x_2)})  \: d\eta 
 \displaystyle = 0, \:\:\: x_2 > d,
\end{aligned}
\label{eq49}
\ee
\be
\begin{aligned}
\displaystyle  \int_0^\infty  \chi_1 \: M_1(\eta) \: \text{Re} (e^{(i \eta x_2)})  \: d\eta 
 \displaystyle = \frac{D_0}{\overline{\kappa^{(u)}_{11}}}, \:\:\: 0<x_2< d,
\end{aligned}
\label{eq50}
\ee
\be
\int_0^\infty \left[ \left(M_1(\eta)+\frac{\overline{e^{(u)}_{15}}}{\overline{\kappa^{(u)}_{11}}} K_1(\eta) \right) -\left(M_2(\eta)+\frac{e^{(l)}_{15}}{\kappa^{(l)}_{11}} K_2(\eta) \right) \right]\text{Re} (e^{(i \eta x_2)})  d\eta =0,  \:\:\: x_2 > d,
\label{eq51}
\ee
and
\be
\begin{aligned}
\displaystyle  \int_0^\infty  \lambda_1 \:L_1 (\eta ) \: \text{Im} (e^{(i \eta x_2)}\: \: d\eta 
 \displaystyle = 0 , \:\:\: 0<x_2< d,
\end{aligned}
\label{eq52}
\ee
\be
\begin{aligned}
\displaystyle  \int_0^\infty  [L_1 (\eta) -L_2 (\eta )] \: \text{Im} (e^{(i \eta x_2)})\:  \: d\eta 
 \displaystyle = 0, \:\:\: x_2 > d,
\end{aligned}
\label{eq53}
\ee
\be
\begin{aligned}
\displaystyle  \int_0^\infty  \chi_1 \:N_1 (\eta ) \: \text{Im} (e^{(i \eta x_2)}\:  \: d\eta 
 \displaystyle = 0, \:\:\: 0<x_2< d,
\end{aligned}
\label{eq54}
\ee
\be
\int_0^\infty \left[ \left(N_1(\eta)+\frac{\overline{e^{(u)}_{15}}}{\overline{\kappa^{(u)}_{11}}} L_1(\eta) \right) -\left(N_2(\eta)+\frac{e^{(l)}_{15}}{\kappa^{(l)}_{11}} L_2(\eta) \right) \right] \text{Im} (e^{(i \eta x_2)})  d\eta =0,  \:\:\: x_2 > d.
\label{eq55}
\ee
The following assumptions are employed to solve equations (\ref{eq48})–(\ref{eq51}):

\begin{equation}
\begin{bmatrix}
g_1(\eta) \\
g_2(\eta)
\end{bmatrix}
=
\begin{bmatrix}
1 & -1 \\
\frac{\overline{e^{(u)}_{15}}}{\overline{\kappa^{(u)}_{11}}} & -\frac{e^{(l)}_{15}}{\kappa^{(l)}_{11}}
\end{bmatrix}
\begin{bmatrix}
K_1(\eta) \\
K_2(\eta)
\end{bmatrix}
+
\begin{bmatrix}
0 \\
M_1(\eta) - M_2(\eta)
\end{bmatrix}
\label{eq56}
\end{equation}
Incorporating the boundary conditions given in (\ref{bc1})-(\ref{bc4}) along with Eqs. (\ref{eq56}), leads to the following results: 
\begin{align}
K_1=\frac{-\left( m_3\frac{e^{(l)}_{15}}{\kappa^{(l)}_{11}}+m_1 m_4\right) }{m_3 n_1- m_2 m_1}g_1+\frac{m_3}{m_3 n_1-m_1 m_2} g_2,  \:\:\:
M_1=\frac{m_4 n_1+\frac{e^{(l)}_{15}}{\kappa^{(l)}_{11}}m_2}{m_3 n_1-m_1 m_2} g_1 - \frac{m_2}{m_3 n_1-m_1 m_2} g_2, \label{eq5958}
\end{align}
where,
\begin{align*}
    m_1=\frac{\overline{\kappa^{(u)}_{11}} \chi_1}{\kappa^{(l)}_{11} \eta}+1,\:\:\:
    m_2={\beta^{(u)}_f} \lambda_1+{\beta^{(l)}_f} \gamma_2, \:\:\:
    m_3=n_2 \chi_1, \:\:\:
    m_4=\beta^{(l)}_f \gamma_2, \:\:\: n_1=\frac{\overline{e^{(u)}_{15}}}{\overline{\kappa^{(u)}_{11}}}-\frac{e^{(l)}_{15}}{\kappa^{(l)}_{11}} , \:\:\:
    n_2=\overline{e^{(u)}_{15}}-\frac{\overline{\kappa^{(u)}_{11}} }{\kappa^{(l)}_{11}}{e^{(l)}_{15}}.
    \label{eq61}
\end{align*}
The results from substituting $K_1$ and $M_1$ into Eqs. (\ref{eq56}) in conjunction with Eqs. (\ref{eq48})-(\ref{eq51}) are as follows:
\be
\ba{lll}
\begin{aligned}
    &K_0\int_0^\infty \eta \: g_1(\eta) \: \text{Re} (e^{(i \eta x_2)}) \: d \eta + \int_0^\infty  g_1(\eta) \: I_1(\eta ) \: \text{Re} (e^{(i \eta x_2)}) \:  d \eta  \\
&+\int_0^\infty \frac{m_3 \lambda_1}{m_3 n_1-m_2 m_1} g_2(\eta) \:  \text{Re} (e^{(i \eta x_2)}) \: d\eta = \frac{1}{\beta^{(u)}_f} \:\left[ T_0-\frac{\overline{e^{(u)}_{15}}}{\overline{\kappa^{(u)}_{11}}}  D_0\right], \:\:\: 0<x_2< d,
\end{aligned}
\ea
\label{eq62}
\ee

\be
\begin{aligned}
\displaystyle  \int_0^\infty   \:g_1(\eta
)\: \text{Re} (e^{(i \eta x_2)})  \: d\eta 
 \displaystyle = 0, \:\:\: x_2 > d,
\end{aligned}
\label{eq63}
\ee

\be
\ba{lll}
\begin{aligned}
     &L_0\int_0^\infty \eta \: g_2(\eta) \: \text{Re} (e^{(i \eta x_2)}) \: d \eta +M_0\int_0^\infty \eta \: g_1(\eta) \: \text{Re} (e^{(i \eta x_2)}) \: d \eta \\
    &
    + \int_0^\infty g_1(\eta) \left[ \:  \frac{\chi_1\left( m_4 n_1+m_2 \: \frac{{e^{(l)}_{15}}}{{\kappa^{(l)}_{11}}} \right)}{m_3 n_1-m_1 m_2}- M_0 \eta  \:\right]  \text{Re} (e^{(i \eta x_2)}) \: d\eta \\
    &+ \int_0^\infty g_2(\eta) \: I_2(\eta)\: \text{Re} (e^{(i \eta x_2)}) \: d \eta =\frac{D_0}{\overline{\kappa^{(u)}_{11}}}, \:\:\: 0<x_2< d,
\end{aligned}
\ea
\label{eq64}
\ee

\be
\int_0^\infty \: g_2(\eta) \:\text{Re} (e^{(i \eta x_2)})  d\eta =0,  \:\:\: x_2 > d,
\label{eq65}
\ee
where,
\begin{equation*}
\ba{lll}
\displaystyle I_1(\eta) = \frac{-\left( m_3\frac{e^{(l)}_{15}}{\kappa^{(l)}_{11}}+m_1 m_4\right) \lambda_1 }{m_3 n_1 - m_2 m_1} - \eta K_0, \:\:\:
I_2(\eta) = \frac{m_2 \chi_1}{m_3 n_1 - m_1 m_2} - \eta L_o, \\ [2ex]
\displaystyle K_0 = -\frac{n_2 \frac{e^{(l)}_{15}}{\kappa^{(l)}_{11}} + \left( \frac{\overline{\kappa^{(u)}_{11}}}{{\kappa^{(l)}_{11}}} + 1 \right) \beta^{(l)}_f m_5}{n_1 n_2 - (\beta^{(u)}_f m_5 + \beta^{(l)}_f m_6)   \left( \frac{\overline{\kappa^{(u)}_{11}}}{{\kappa^{(l)}_{11}}} + 1 \right)}, \:\:\:
L_0=\frac{\beta^{(u)}_f m_5 + \beta^{(l)}_f m_6 }{n_1 n_2 - (\beta^{(u)}_f m_5 + \beta^{(l)}_f m_6)   \left( \frac{\overline{\kappa^{(u)}_{11}}}{\kappa^{(l)}_{11}} + 1 \right)},\\ [2ex]
\displaystyle M_0=\frac{\beta^{(l)}_f m_6+\frac{e^{(l)}_{15}}{\kappa^{(l)}_{11}} \: (\beta^{(u)}_f m_5 + \beta^{(l)}_f m_6)}{n_1 n_2 - (\beta^{(u)}_f m_5 + \beta^{(l)}_f m_6)   \left( \frac{\overline{\kappa^{(u)}_{11}}}{\kappa^{(l)}_{11}} + 1 \right)}, \:\:\:
m_5=\sqrt{\left( 1 + \frac{{\Im^{(u)}}^{'}}{\beta^{(u)}_f} - \frac{c^2}{{\beta^{(u)}}^2} \right)}, \:\:\:  \:\:\: m_6=\sqrt{\left(1-\frac{c^2}{{\beta^{(l)}}^2} \right). }
\ea
\label{I1I2KLM}
\end{equation*}
 The solution of the integral equations (\ref{eq62})-(\ref{eq65}) are expressed in the following form:
 \begin{align}
\begin{bmatrix}
g_1(\eta) \\
g_2(\eta)
\end{bmatrix}
&=
\begin{bmatrix}
\displaystyle \frac{1}{K_0} \int_0^d \alpha_1(t) \: J_0(t \eta) \: dt \\
\displaystyle \frac{1}{L_0} \int_0^d \beta_1(t) \: J_0(t \eta) \: dt
\end{bmatrix},
\label{eqg2}
\end{align} 
where \( J_v(\cdot) \) (\( v \geq 0 \)) represents the Bessel function of the first kind, and \( \alpha_1(t) \) and \( \beta_1(t) \) are the functions to be evaluated from the two double integral equations.  Consequently, the solutions to these double integral equations (\ref{eq62}) through (\ref{eq65})  are formulated as:
\be
\alpha_1(t)+t \int_0^d \alpha_1(s) \: G_1(s,t)ds+t \int_0^d \beta_1(s) \:  G_2(s,t)ds =\frac{t}{\beta^{(u)}_f} \:\left[ T_0-\frac{\overline{e^{(u)}_{15}}}{\overline{\kappa^{(u)}_{11}}}  D_0\right], \:\:\: 0<x_2< d,
\label{eq73}
\ee
\be
\frac{\alpha_1(t) \: M_0}{K_0} + \beta_1(t)+ t\int_0^d \beta_1(s) \: G_3(s,t)ds +t\int_0^d \alpha_1(s) \: G_4(s,t)ds=\frac{D_0 \: t}{\overline{\kappa^{(u)}_{11}}}, \:\:\: 0<x_2< d.
\label{eq74}
\ee
Equations (\ref{eq73}) and (\ref{eq74}) represent a coupled system of  simultaneous Fredholm integral equations, and the functions $G_i (s,t)$, where $i=1,2,3,4$ are described by:
%%%%%%%%%% summation form===================================
\begin{equation}
\begin{aligned}
G_1(s,t) &= \frac{1}{K_0} \sum_{i=1}^N I_1(\eta_i) \: J_0(s \eta_i) \: J_0(t \eta_i) \Delta \eta, \\
G_2(s,t) &= \frac{1}{L_0} \sum_{i=1}^N \frac{m_3 \lambda_1}{m_3 n_1 - m_1 m_2} \: J_0(s \eta_i) \: J_0(t \eta_i) \Delta \eta, \\
G_3(s,t) &= \frac{1}{L_0} \sum_{i=1}^N I_2(\eta_i) \: J_0(s \eta_i) \: J_0(t \eta_i) \Delta \eta, \\
G_4(s,t) &= \frac{1}{K_0} \sum_{i=1}^N \left[ \frac{\chi_1 \left( m_4 n_1 + m_2 \frac{e^{(l)}_{15}}{\kappa^{(l)}_{11}} \right)}{m_3 n_1 - m_1 m_2} - M_0 \eta_i \right] J_0(s \eta_i) \: J_0(t \eta_i) \Delta \eta,
\end{aligned}
\label{eq75-78}
\end{equation}
To solve Eqs. (\ref{eq52})–(\ref{eq55}), consider the following system::
\begin{equation}
\begin{bmatrix}
g_3(\eta) \\
g_4(\eta)
\end{bmatrix}
=
\begin{bmatrix}
1 & -1 \\
\frac{\overline{e^{(u)}_{15}}}{\overline{\kappa^{(u)}_{11}}} & -\frac{e^{(l)}_{15}}{\kappa^{(l)}_{11}}
\end{bmatrix}
\begin{bmatrix}
L_1(\eta) \\
L_2(\eta)
\end{bmatrix}
+
\begin{bmatrix}
0 \\
N_1(\eta) - N_2(\eta)
\end{bmatrix}.
\label{eq79}
\end{equation}
Using boundary conditions (\ref{bc1})-(\ref{bc4}) along with expression (\ref{eq79}), the following expressions are derived:
\begin{align}
L_1=\frac{-\left( m_3\frac{e^{(l)}_{15}}{\kappa^{(l)}_{11}}+m_1 m_4\right) }{m_3 n_1- m_2 m_1}g_3+\frac{m_3}{m_3 n_1-m_1 m_2} g_4, \:\:\: 
N_1=\frac{m_4 n_1+\frac{e^{(l)}_{15}}{\kappa^{(l)}_{11}}m_2}{m_3 n_1-m_1 m_2} g_3 - \frac{m_2}{m_3 n_1-m_1 m_2} g_4. \label{eq8281}
\end{align}
By substituting $L_1$ and $N_1$ into Eqs. (\ref{eq52})-(\ref{eq55}) and subsequently applying Eq. (\ref{eq79}), the resulting forms of Eqs. (\ref{eq52})-(\ref{eq55}) are:
\be
\ba{lll}
\begin{aligned}
    &K_0\int_0^\infty \eta \: g_3(\eta) \: \text{Im} (e^{(i \eta x_2)}) \: d \eta + \int_0^\infty  g_3(\eta) \: I_1(\eta ) \: \text{Im} (e^{(i \eta x_2)}) \:  d \eta  \\
&+\int_0^\infty \frac{m_3 \lambda_1}{m_3 n_1-m_2 m_1} g_4(\eta) \:  \text{Im} (e^{(i \eta x_2)}) \: d\eta =0, \:\:\: 0<x_2< d,
\end{aligned}
\ea
\label{eq83}
\ee

\be
\begin{aligned}
\displaystyle  \int_0^\infty   \:g_3(\eta
)\: \text{Im} (e^{(i \eta x_2)})  \: d\eta 
 \displaystyle = 0, \:\:\: x_2 > d,
\end{aligned}
\label{eq84}
\ee

\be
\ba{lll}
\begin{aligned}
     &L_0\int_0^\infty \eta \: g_4(\eta) \: \text{Im} (e^{(i \eta x_2)}) \: d \eta +M_0\int_0^\infty \eta \: g_3(\eta) \: \text{Im} (e^{(i \eta x_2)}) \: d \eta \\
    &
    + \int_0^\infty g_3(\eta) \left[ \:  \frac{\chi_1\left( m_4 n_1+m_2 \: \frac{{e^{(l)}_{15}}}{{\kappa^{(l)}_{11}}} \right)}{m_3 n_1-m_1 m_2}- M_0 \eta  \:\right]  \text{Im} (e^{(i \eta x_2)}) \: d\eta \\
    &+ \int_0^\infty g_4(\eta) \: I_2(\eta)\: \text{Im} (e^{(i \eta x_2)}) \: d \eta =0, \:\:\: 0<x_2< d.
\end{aligned}
\ea
\label{eq85}
\ee

\be
\int_0^\infty \: g_4(\eta) \: \text{Im} (e^{(i \eta x_2)})  d\eta =0,  \:\:\: x_2 > d.
\label{eq86}
\ee
The solution of the integral Eqs. (\ref{eq83})-(\ref{eq86}) can be expressed in the following form:
\begin{equation}
\begin{bmatrix}
g_3(\eta) \\
g_4(\eta)
\end{bmatrix}
=
\begin{bmatrix}
\displaystyle \frac{1}{K_0} \int_0^d t \: \alpha_2(t) \: J_1(t \eta) \: dt \\
\displaystyle \frac{1}{L_0} \int_0^d t \: \beta_2(t) \: J_1(t \eta) \: dt
\end{bmatrix},
\label{eq87}
\end{equation}
where, \( \alpha_2(t) \) and \( \beta_2(t) \) satisfy the system of Fredholm integral equations given by:
\be
\alpha_2(t)+ \int_0^d s \: \alpha_2(s) \: H_1(s,t) \: ds+ \int_0^d  s \:  \beta_2(s) \:  H_2(s,t) \: ds =0, \:\:\: 0<x_2< d,
\label{eq89}
\ee
\be
\frac{\alpha_2(t) \: M_0}{K_0} + \beta_2(t)+ \int_0^d s \: \beta_2(s) \: H_3(s,t) \: ds +\int_0^d s \: \alpha_2(s) \: H_4(s,t) \: ds=0, \:\:\: 0<x_2< d,
\label{eq90}
\ee
where,
\begin{equation*}
\begin{aligned}
H_1(s,t) &= \frac{1}{K_0} \sum_{i=1}^N I_1(\eta_i) \: J_1(s \eta_i) \: J_1(t \eta_i) \Delta \eta, \\
H_2(s,t) &= \frac{1}{L_0} \sum_{i=1}^N \frac{m_3 \lambda_1}{m_3 n_1 - m_1 m_2} \: J_1(s \eta_i) \: J_1(t \eta_i) \Delta \eta, \\
H_3(s,t) &= \frac{1}{L_0} \sum_{i=1}^N I_2(\eta_i) \: J_1(s \eta_i) \: J_1(t \eta_i) \Delta \eta, \\
H_4(s,t) &= \frac{1}{K_0} \sum_{i=1}^N \left[ \frac{\chi_1 \left( m_4 n_1 + m_2 \frac{e^{(l)}_{15}}{\kappa^{(l)}_{11}} \right)}{m_3 n_1 - m_1 m_2} - M_0 \eta_i \right] \: J_1(s \eta_i) \: J_1(t \eta_i) \Delta \eta.
\end{aligned}
\label{eq86-89}
\end{equation*}
Solving Eqs. (\ref{eq73}), (\ref{eq74}), (\ref{eq89}), and (\ref{eq90}) numerically yields the solutions for \( \alpha_1(t) \), \( \alpha_2(t) \), \( \beta_1(t) \), and \( \beta_2(t) \) which can then be used in conjunction with Eqs. (\ref{eq25}), (\ref{eq5958}), (\ref{eqg2}), (\ref{eq8281}), and (\ref{eq87}) to determine the SIF and the EDIF at $x_2=d$. These equations provide the necessary relationships to calculate the values of the SIF and the EDIF at the specified location $x_2=d$, given by: 
\begin{align}
K^{d}_{III}=\lim_{x_2\to d^{+}} \left[  \sqrt{2 \pi (x_2-d)} \: \{ \:T^{(u)}_{13}(0,\: x_2) \: \}_{x_2>d} \right]  = \sqrt{\frac{\pi}{d}} \left[ \mathbf{v}^T \mathbf{A} \mathbf{x} \right] ,
\label{eq95}
\end{align}
\begin{align}
K^{D(d)}_{III}=\lim_{x_2\to d^{+}} \left[  \sqrt{2 \pi (x_2-d)} \: \{ \:D^{(u)}_{1}(0,\: x_2) \: \}_{x_2>d} \right] = \sqrt{\frac{\pi}{d}}  \left[ \mathbf{v^{'}}^T \mathbf{A} \mathbf{x} \right]  \: , 
\label{eq97}
\end{align}
where,
\begin{equation*}
\mathbf{v} = \begin{bmatrix} \beta^{(u)}_f \\  -\overline{e^{(u)}_{15}} \end{bmatrix}, 
\mathbf{v^{'}} = \begin{bmatrix} 0 \\ \overline{\kappa^{(u)}_{11}} \end{bmatrix},\quad
\mathbf{A} = \begin{bmatrix} \alpha_1(d) & \alpha_2(d) \\ \beta_1(d) & \beta_2(d) \end{bmatrix}, \quad
\mathbf{x} = \begin{bmatrix} 1 \\ d \end{bmatrix}.
\end{equation*}
At $x_2=-d$, the SIF ($K^{-d}_{III}$) and EDIF  ($K^{D(-d)}_{III}$) are given by:
%%%%%%%%%%
\begin{align}
K^{-d}_{III}= \lim_{x_2\to -d^{-}} \left[  \sqrt{2 \pi (-x_2-d)} \: \{ \:T^{(u)}_{13}(0,\: x_2) \: \}_{x_2<-d} \right] = \sqrt{\frac{\pi}{d}} \left[ \mathbf{w}^T \mathbf{A} \mathbf{y} \right],
\label{eq96}
\end{align} 
\begin{align}
K^{D(-d)}_{III}&=\lim_{x_2\to -d^{-}} \left[  \sqrt{2 \pi (-x_2-d)} \: \{ \:D^{(u)}_{1}(0,\: x_2) \: \}_{x_2<-d} \right] = \sqrt{\frac{\pi}{d}} \left[ \mathbf{w^{'}}^T \mathbf{A} \mathbf{y} \right],
\label{eq98}
\end{align}
where,
\begin{equation*}
\mathbf{A} = \begin{bmatrix} \alpha_1(d) & \alpha_2(d) \\ \beta_1(d) & \beta_2(d) \end{bmatrix}, \quad
\mathbf{w} = \begin{bmatrix} -\beta^{(u)}_f \\ \overline{e^{(u)}_{15}} \end{bmatrix}
, \quad
\mathbf{w^{'}} = \begin{bmatrix} 0 \\ -\overline{\kappa^{(u)}_{11}} \end{bmatrix}, \quad
\mathbf{y} = \begin{bmatrix} 1 \\ -d \end{bmatrix}.
\end{equation*}
Expressions for the SIF and EDIF near the vicinity of the crack tip on both the right and left sides have been derived, as shown in Eqs. (\ref{eq95}), (\ref{eq97}), (\ref{eq96}), and (\ref{eq98}). This derivation is the main objective of this study. 
In the following section, numerical simulations are used to provide a graphical representation that illustrates the impact of various parameters on these factors.
\section{Numerical calculation and graphical representation }
\label{Numerical Calculation and Graphical Representation}
\begin{table}[htbp]
\centering
\caption{Material constants \cite{singh2019love, chen2024rayleigh}.}
\begin{tabular}{|l |l l l|} 
\hline
\textbf{Material Constant} & \textbf{Epoxy-BNKLBT} & \textbf{Epoxy-KNLNTS} & \textbf{Epoxy-PZT7A} \\

\hline
$\mathbf{{(\mu_{44})_e}}$ (GPa) & 7.511 & 8.172 & 5.895 \\
$\mathbf{(\mu_{44})_v}$ (GPa s) & 9.7252 & 32.239 & 22.226 \\
$\mathbf{({e_{15}})_e}$ (Cm$^{-2}$) & -0.0919 & -0.0968 & 0.6751 \\
$\mathbf{({e_{15}})_v}$ (Csm$^{-2}$) & -2.85$\times 10^{-4}$ & -3$\times 10^{-4}$ & 0.21$\times 10^{-4}$ \\
$\mathbf{({\kappa_{11}})_e}$ (C$^2$ N$^{-1}$ m$^{-2}$) & -0.3041$\times 10^{-9}$ & -0.4187$\times 10^{-9}$ & 2.00026$\times 10^{-9}$ \\
$\mathbf{({\kappa_{11}})_v}$ (C$^2$ s N$^{-1}$ m$^{-2}$) & -0.02$\times 10^{-12}$ & -2.55$\times 10^{-12}$ & -2.9$\times 10^{-12}$ \\
$\mathbf{\rho}$ (kg m$^{-3}$) & 3949 & 3259 & 5426 \\
\hline
\end{tabular}
\label{material_constant}
\end{table}

In this study, numerical simulations are conducted to evaluate the SIF and DEIF for mode-III fractures induced by propagating Love waves in a bimaterial system. The upper half-space is modeled with Epoxy-BNKLBT, while the lower half-space is varied among two materials: Epoxy-KNLNTS, and Epoxy-PZT7A. Equation (\ref{eq95}), which provides the expression for the SIF, and Eq. (\ref{eq97}), which shows the EDIF, are used to analyze the numerical results. This variation allows for the exploration of how different material properties influence the SIF, EDIF, and their relationship with Love wave velocity. The SIF for various combinations of upper and lower half-space materials is computed and plotted to visualize the dependence of the SIF on Love wave velocity.
The material parameters for the two half-spaces are detailed in Table \ref{material_constant}. In the graph, the dotted lines represent the bimaterial I structure composed of Epoxy-BNKLBT and Epoxy-KNLNTS and the bold line indicates the bimaterial II structure composed of Epoxy-BNKLBT and Epoxy-PZT7A.

\bfg[htbp]
\centering
\begin{subfigure}[b] {0.45\textwidth}
\includegraphics[width=\textwidth ]{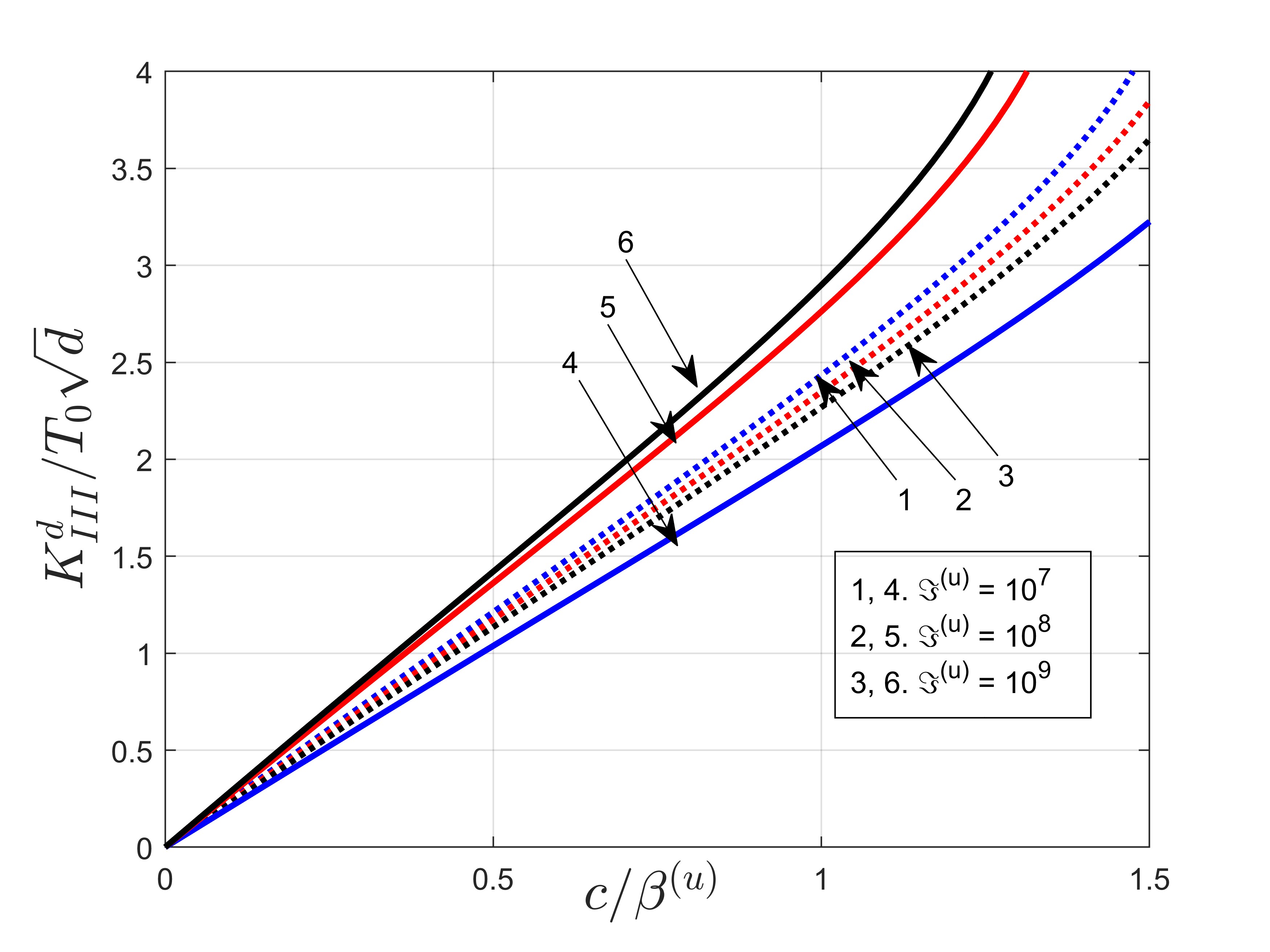}
\caption{}
\label{IS}
\end{subfigure}
~
\begin{subfigure}[b] {0.45\textwidth}
\includegraphics[width=\textwidth ]{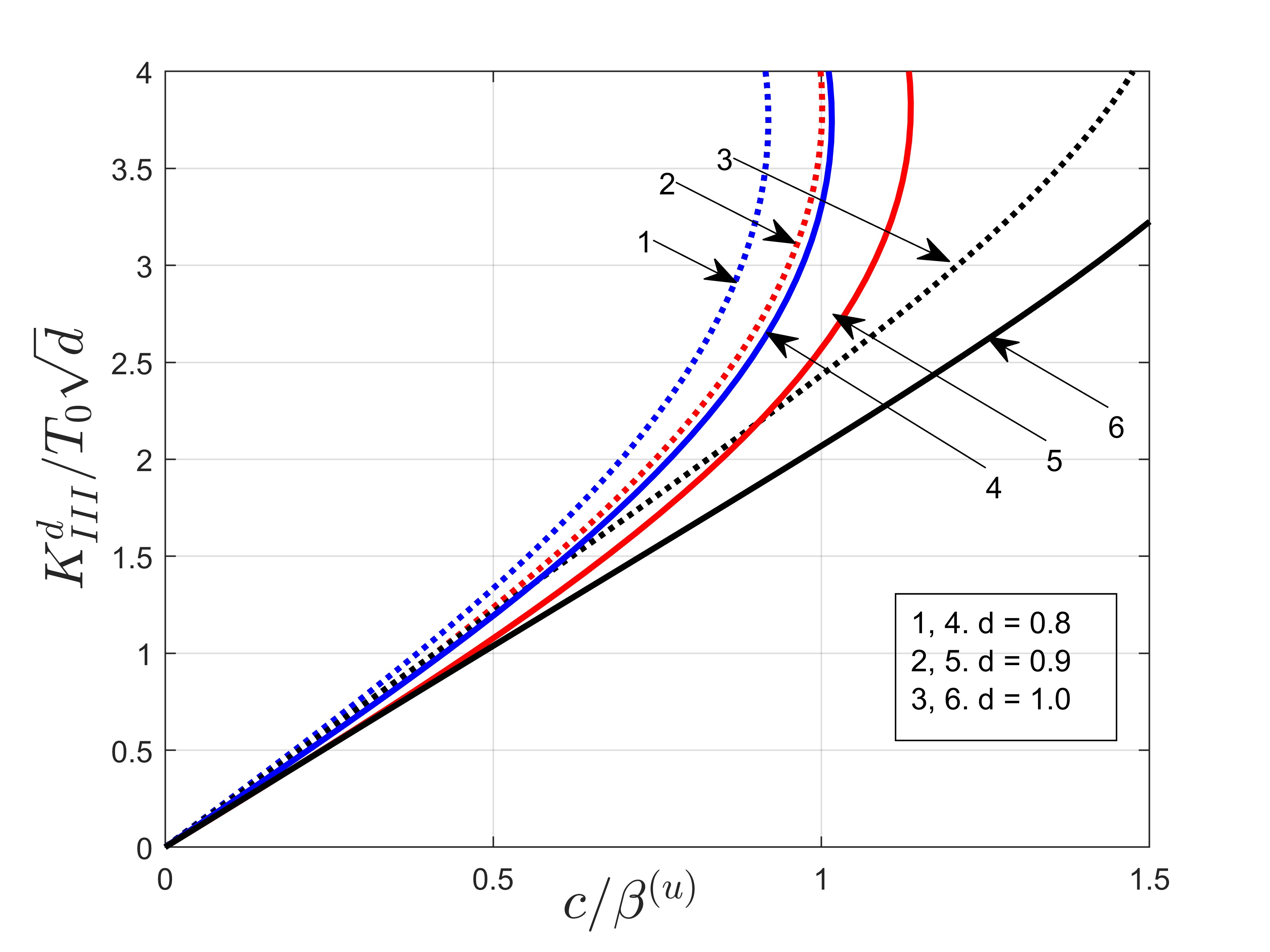}
\caption{}
\label{d}
\end{subfigure}
~
\begin{subfigure}[b] {0.45\textwidth}
\includegraphics[width=\textwidth ]{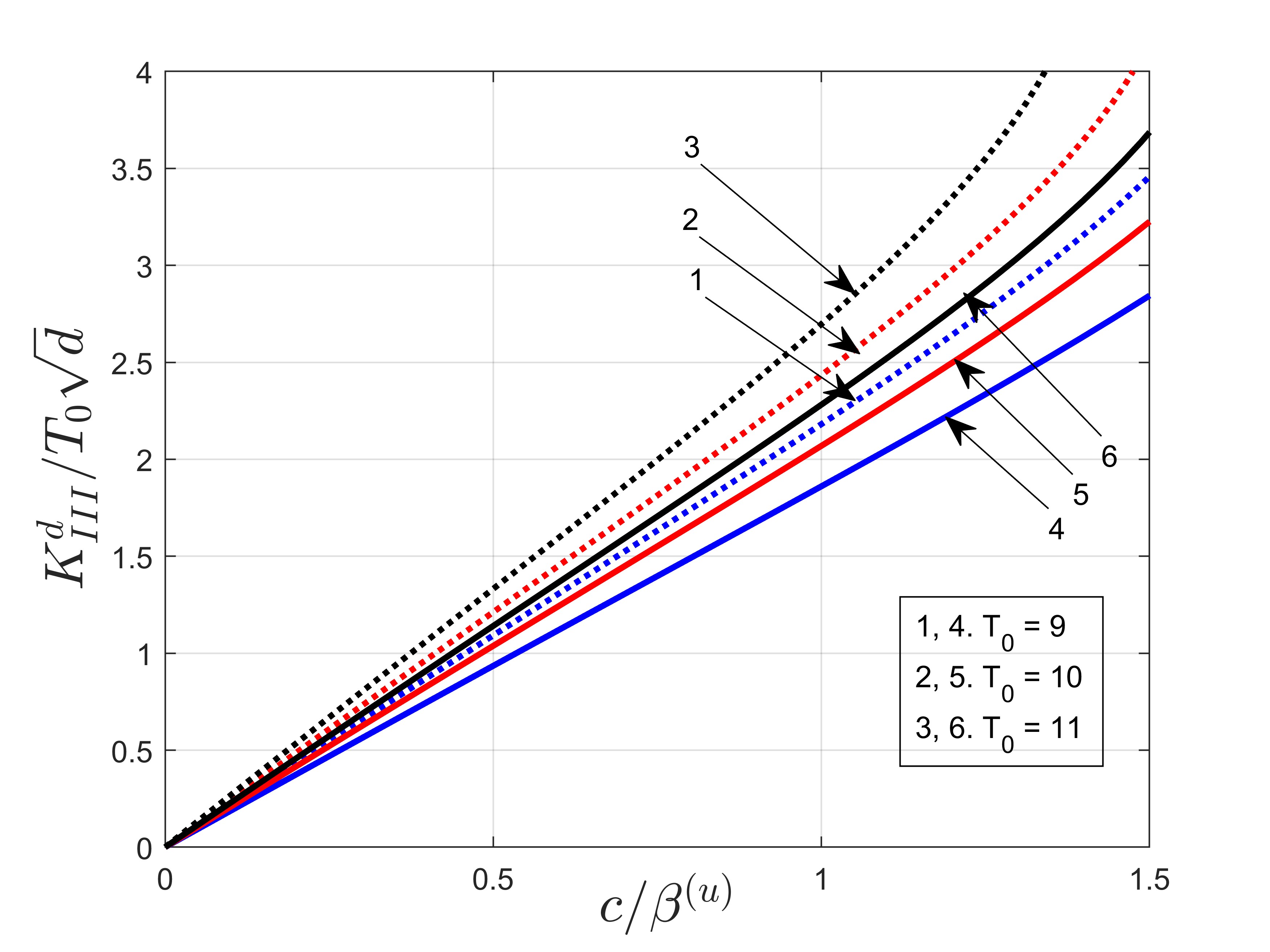}
\caption{}
\label{T0}
\end{subfigure}
~
\begin{subfigure}[b] {0.45\textwidth}
\includegraphics[width=\textwidth ]{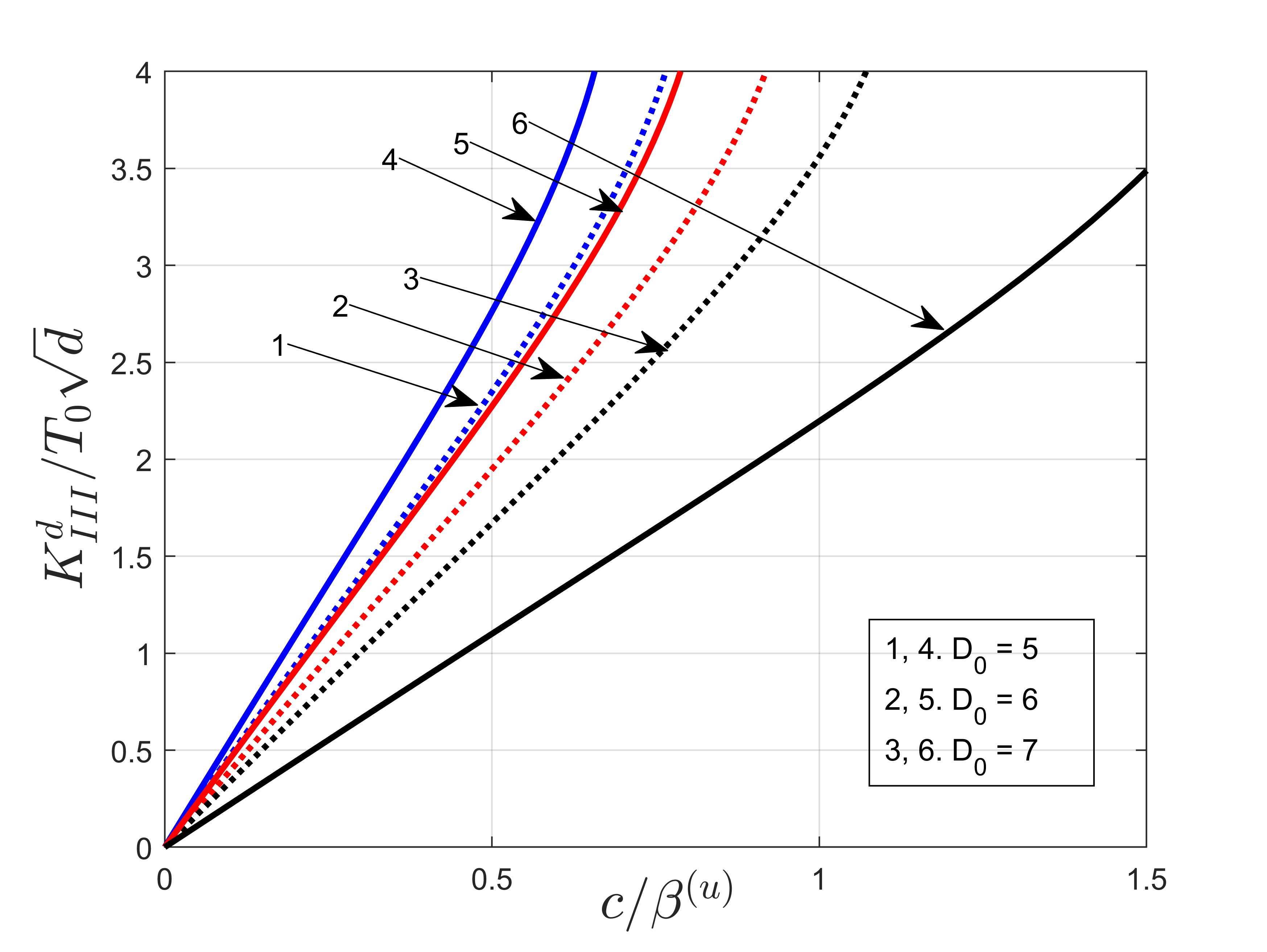}
\caption{}
\label{D0}
\end{subfigure}
~
\begin{subfigure}[b] {0.45\textwidth}
\includegraphics[width=\textwidth ]{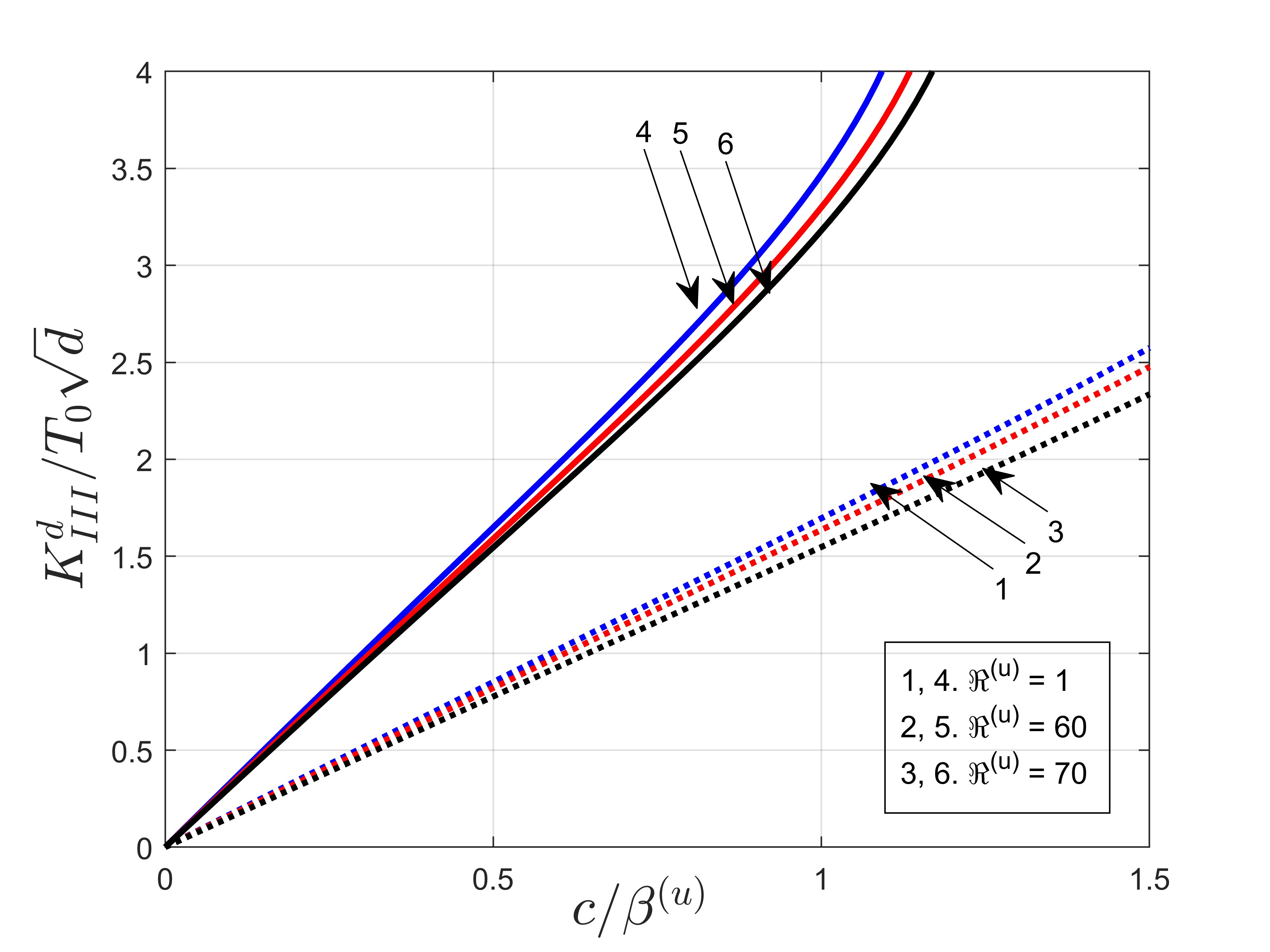}
\caption{}
\label{rotation1}
\end{subfigure}
~
\begin{subfigure}[b] {0.45\textwidth}
\includegraphics[width=\textwidth ]{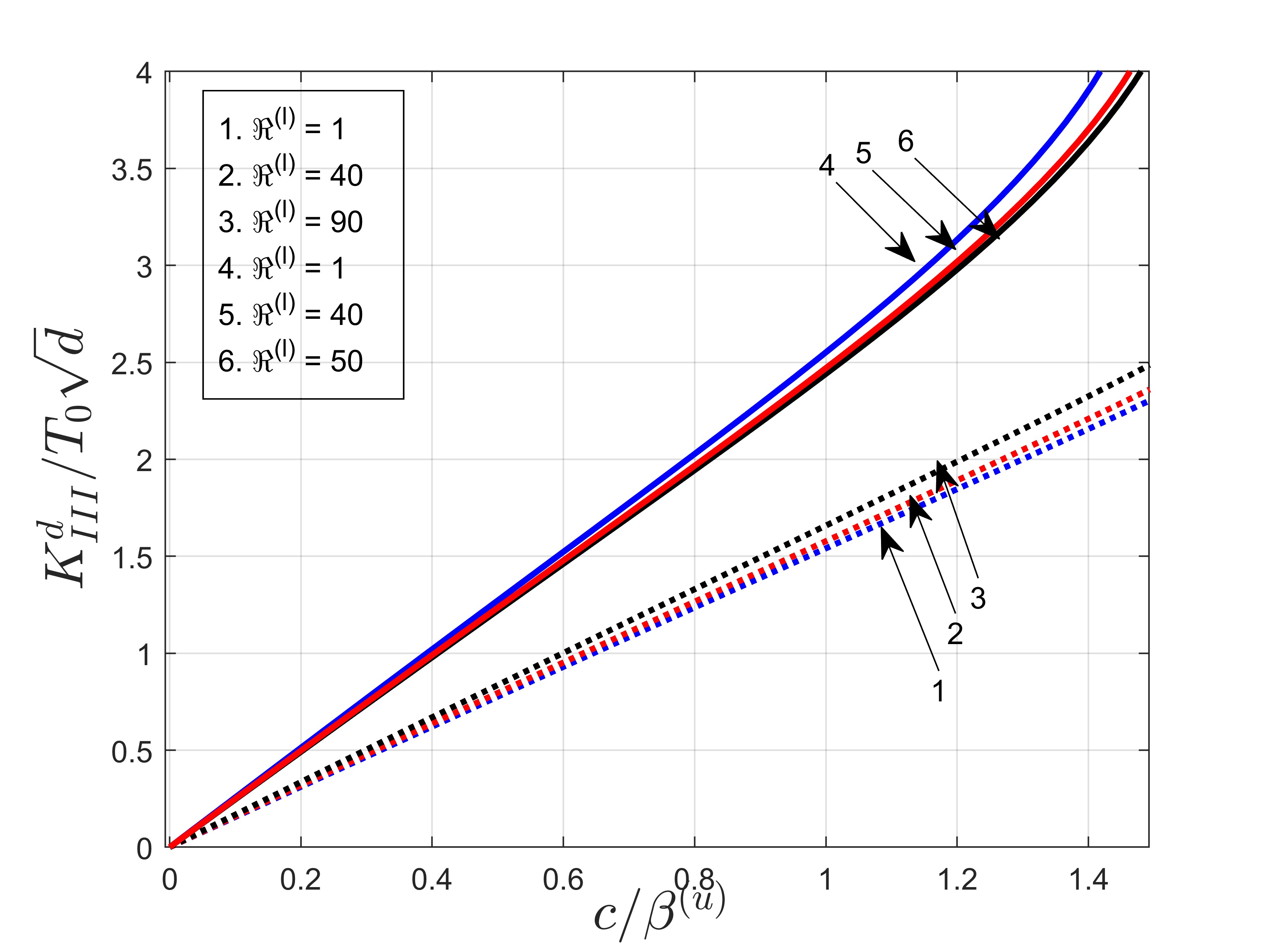}
\caption{}
\label{rotation2}
\end{subfigure}

\caption{Variation of the SIF at the right crack tip as a function of the phase velocity of the Love wave due to variation in (a) initial stress, (b) crack length (c) interface stress (d) interface electric displacement (e) rotation parameter (upper half-space) (f) rotation parameter (lower half-space).}  
\label{variation in IS, d,D T}
\efg
%%%%%%%%%%%%%%%%%%%%%%%%%%%%%%%%%%
\bfg[htbp]
\centering
\begin{subfigure}[b] {0.45\textwidth}
\includegraphics[width=\textwidth ]{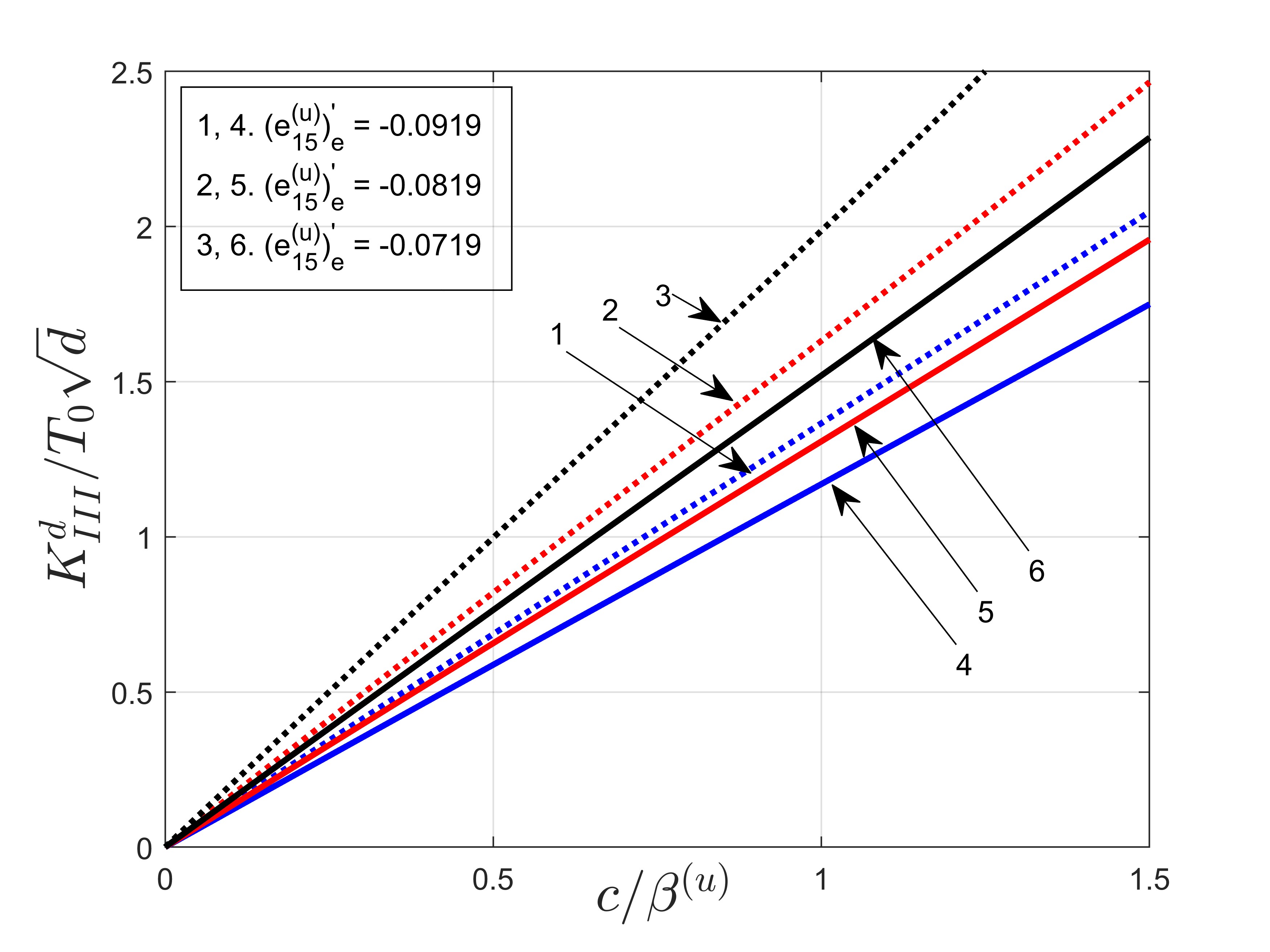}
\caption{}
\label{e15ue}
\end{subfigure}
~
\begin{subfigure}[b] {0.45\textwidth}
\includegraphics[width=\textwidth ]{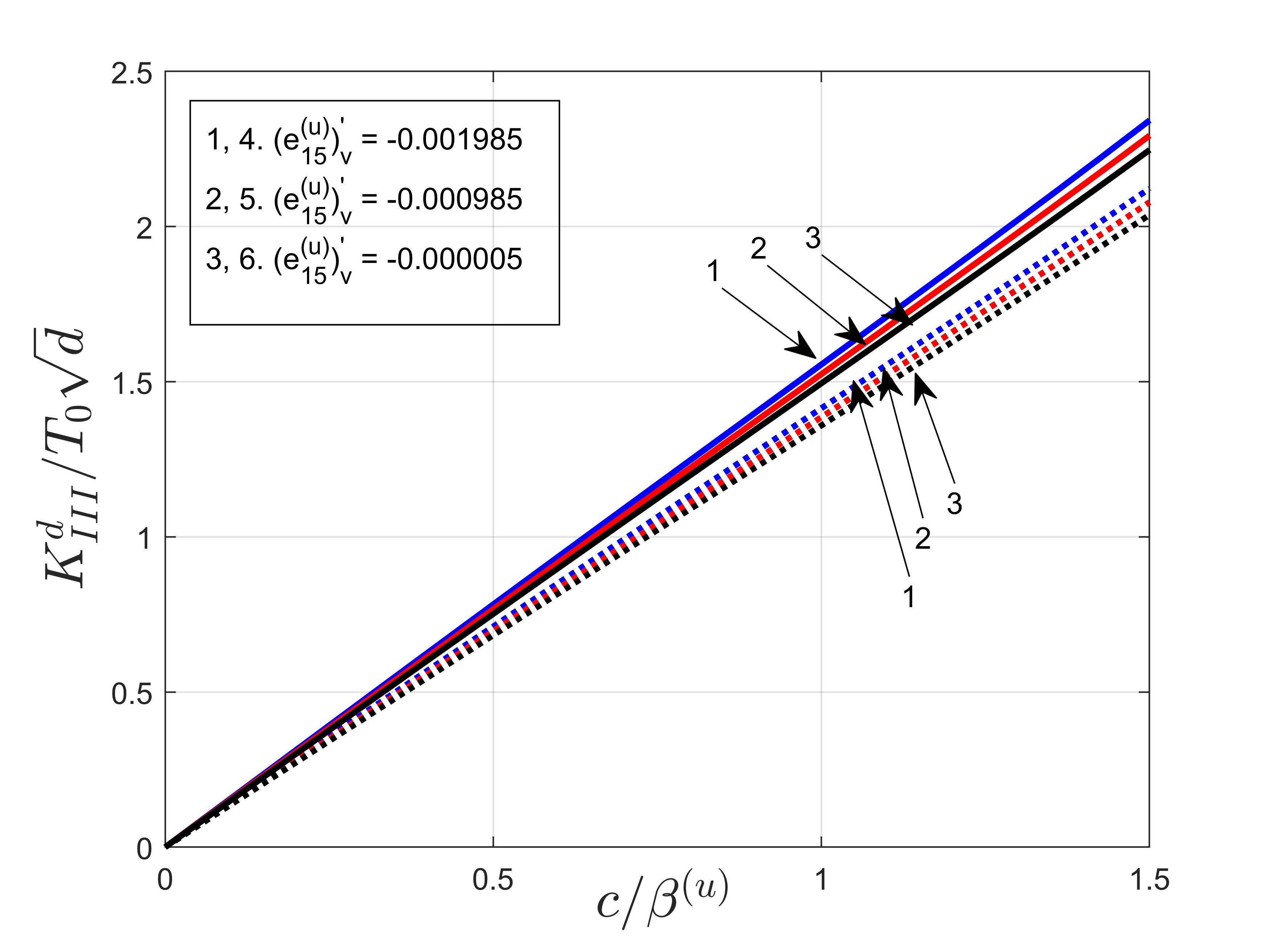}
\caption{}
\label{e15uv}
\end{subfigure}
~
\begin{subfigure}[b] {0.45\textwidth}
\includegraphics[width=\textwidth ]{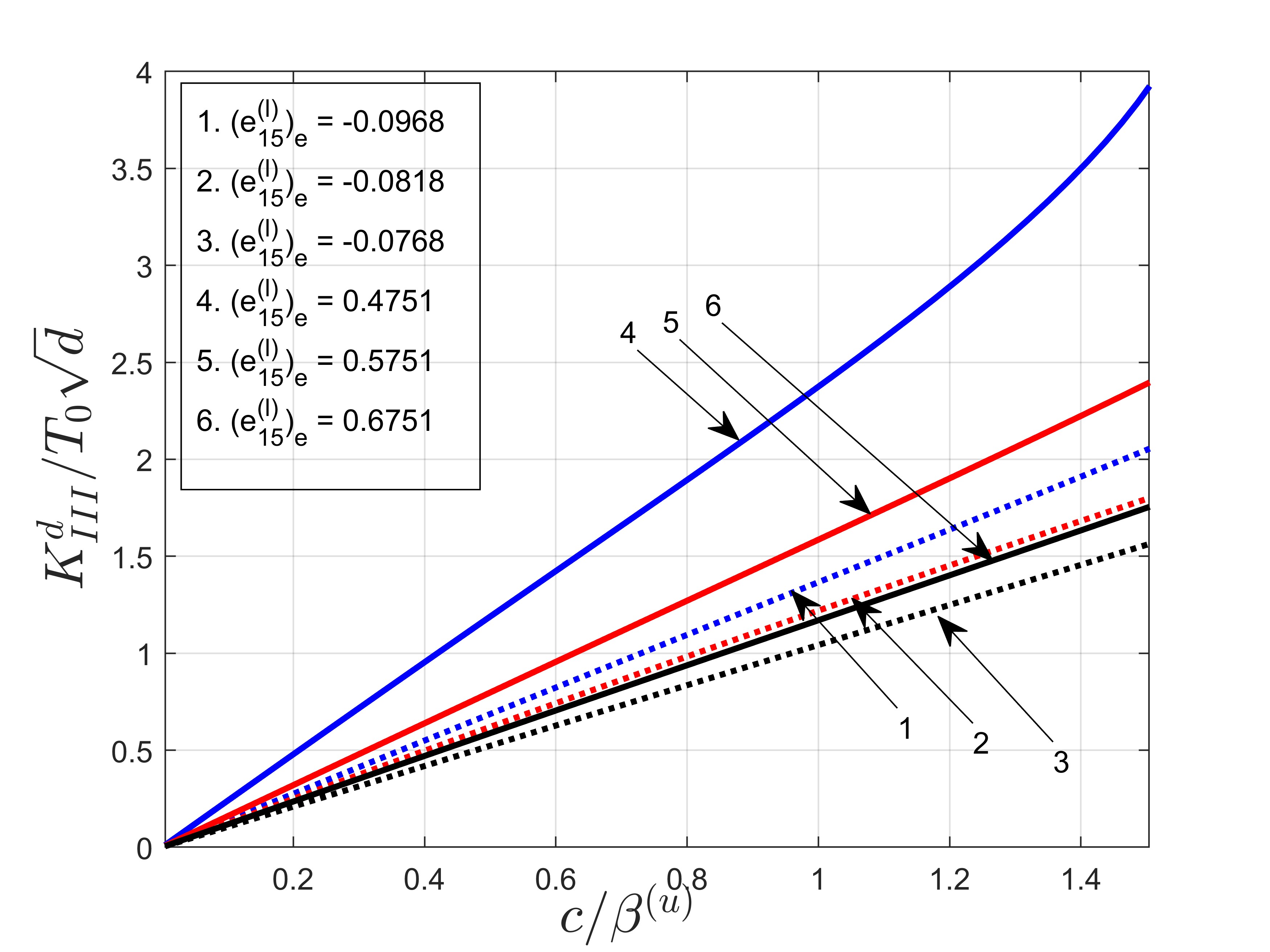}
\caption{}
\label{e15le}
\end{subfigure}
~
\begin{subfigure}[b] {0.45\textwidth}
\includegraphics[width=\textwidth ]{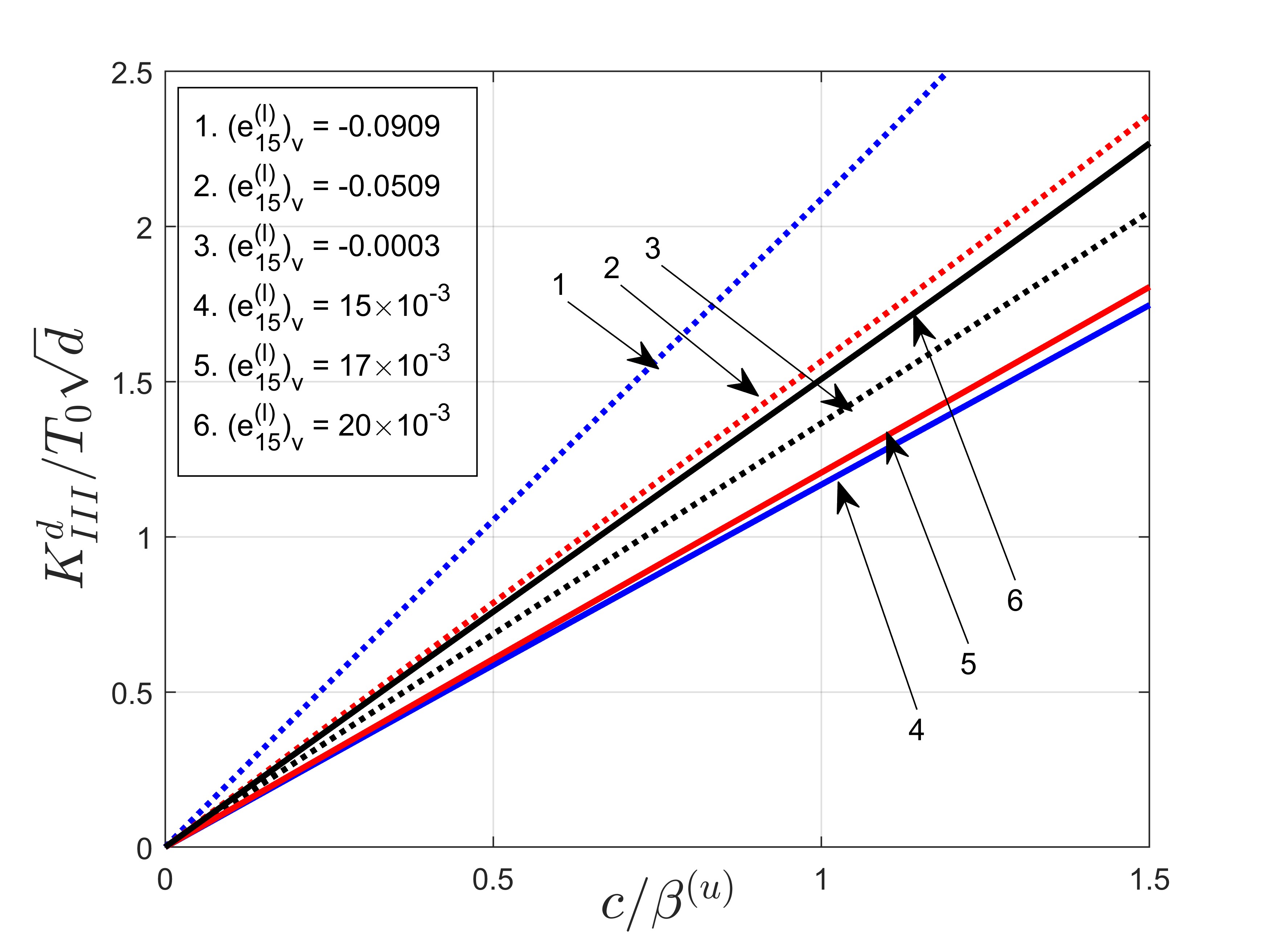}
\caption{}
\label{e15lv}
\end{subfigure}

\caption{Variation of the SIF at the right crack tip as a function of the phase velocity of the Love wave due to variation in (a) piezoelectric constant (upper-half) (b) piezoelectric loss moduli (upper-half) (c) piezoelectric constant (lower-half) (d) piezoelectric loss moduli (lower-half).}
\label{vriation in e15}
\efg
%%%%%%%%%%%%%%%%%%%%%%%%%%%%%%%%%%
\bfg[htbp]
\centering
\begin{subfigure}[b] {0.45\textwidth}
\includegraphics[width=\textwidth ]{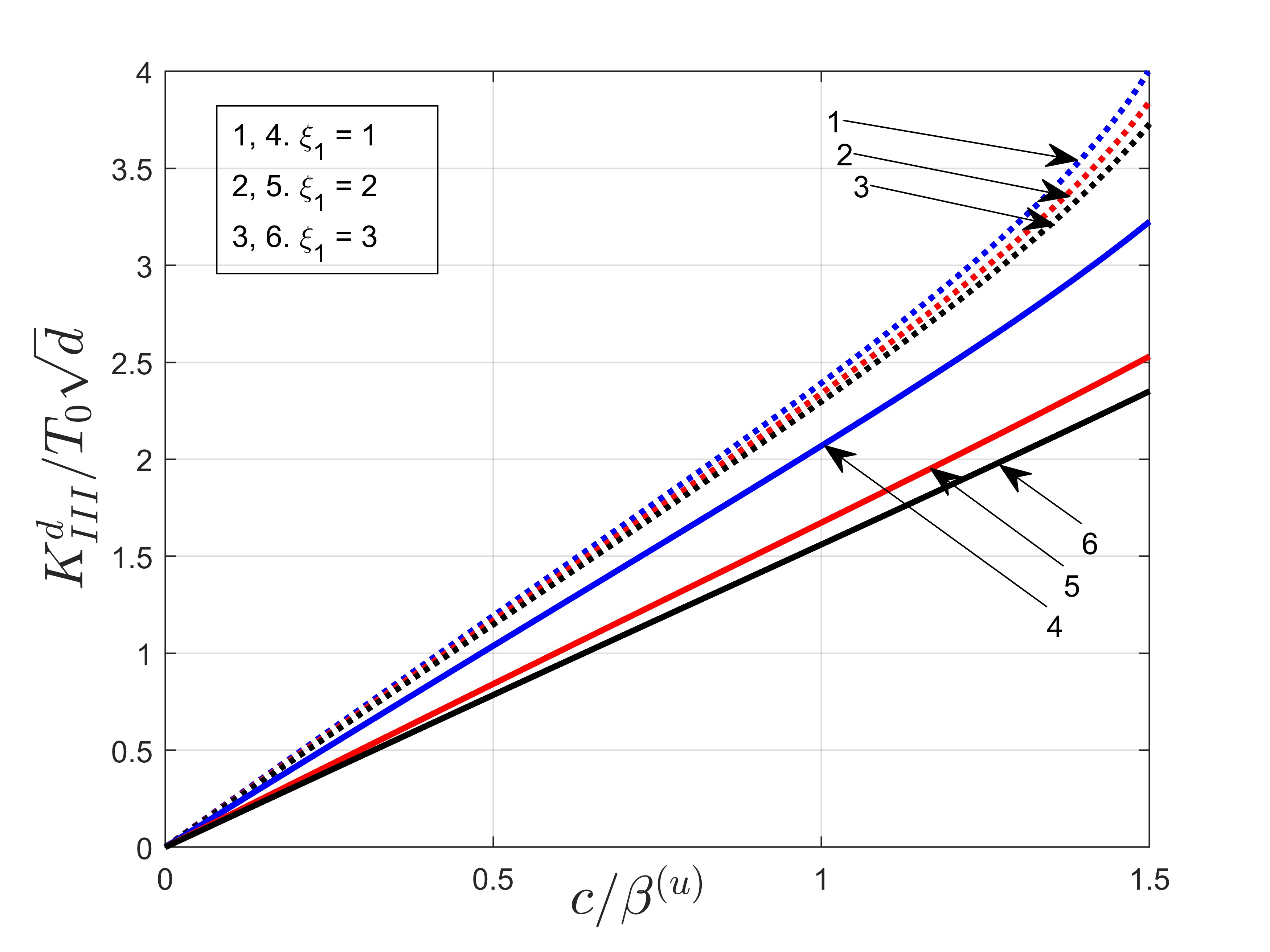}
\caption{}
\label{heterogenity_paramtre}
\end{subfigure}
~
\begin{subfigure}[b] {0.45\textwidth}
\includegraphics[width=\textwidth ]{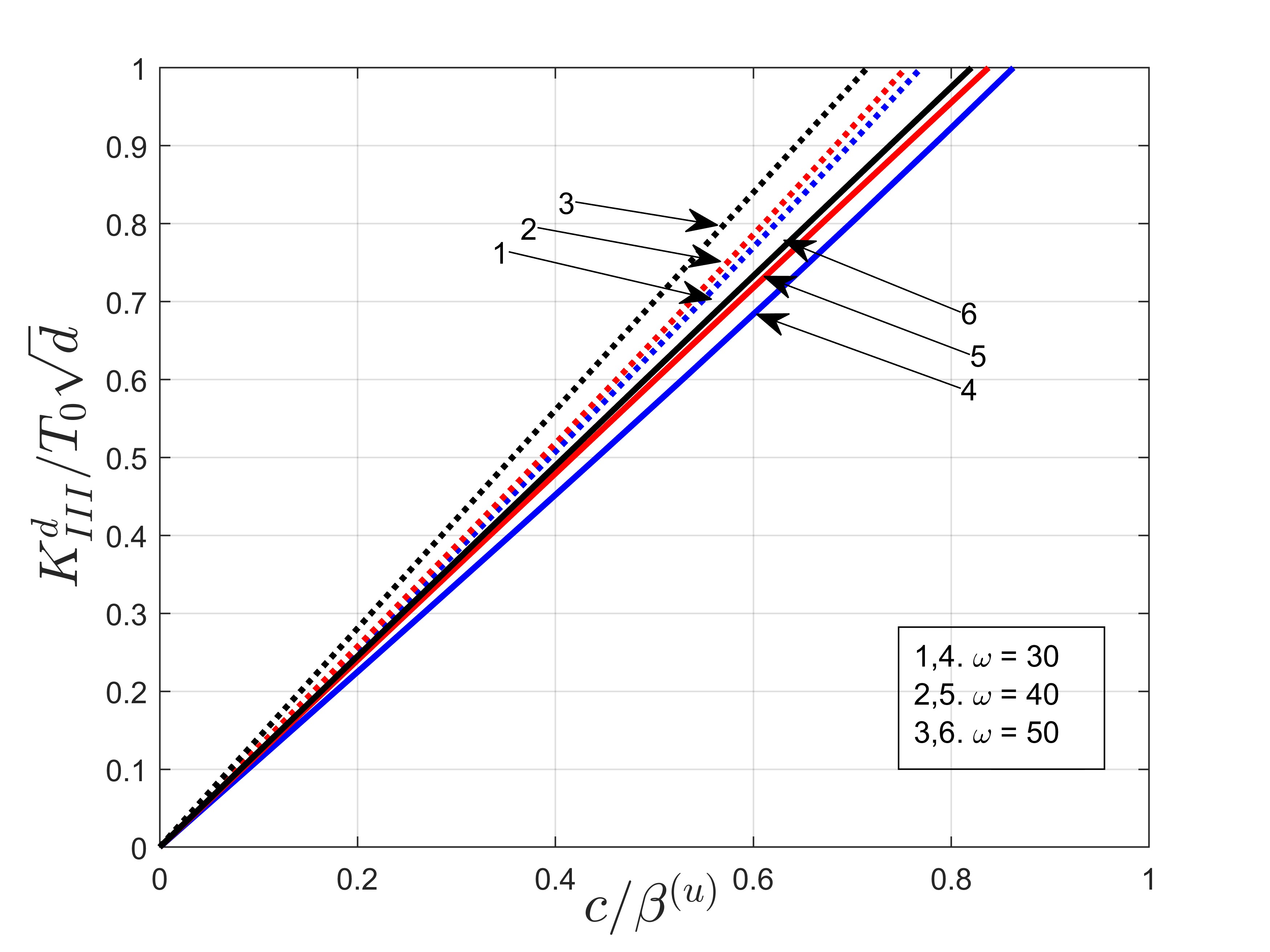}
\caption{}
\label{omegaedited}
\end{subfigure}
~
\begin{subfigure}[b] {0.45\textwidth}
\includegraphics[width=\textwidth ]{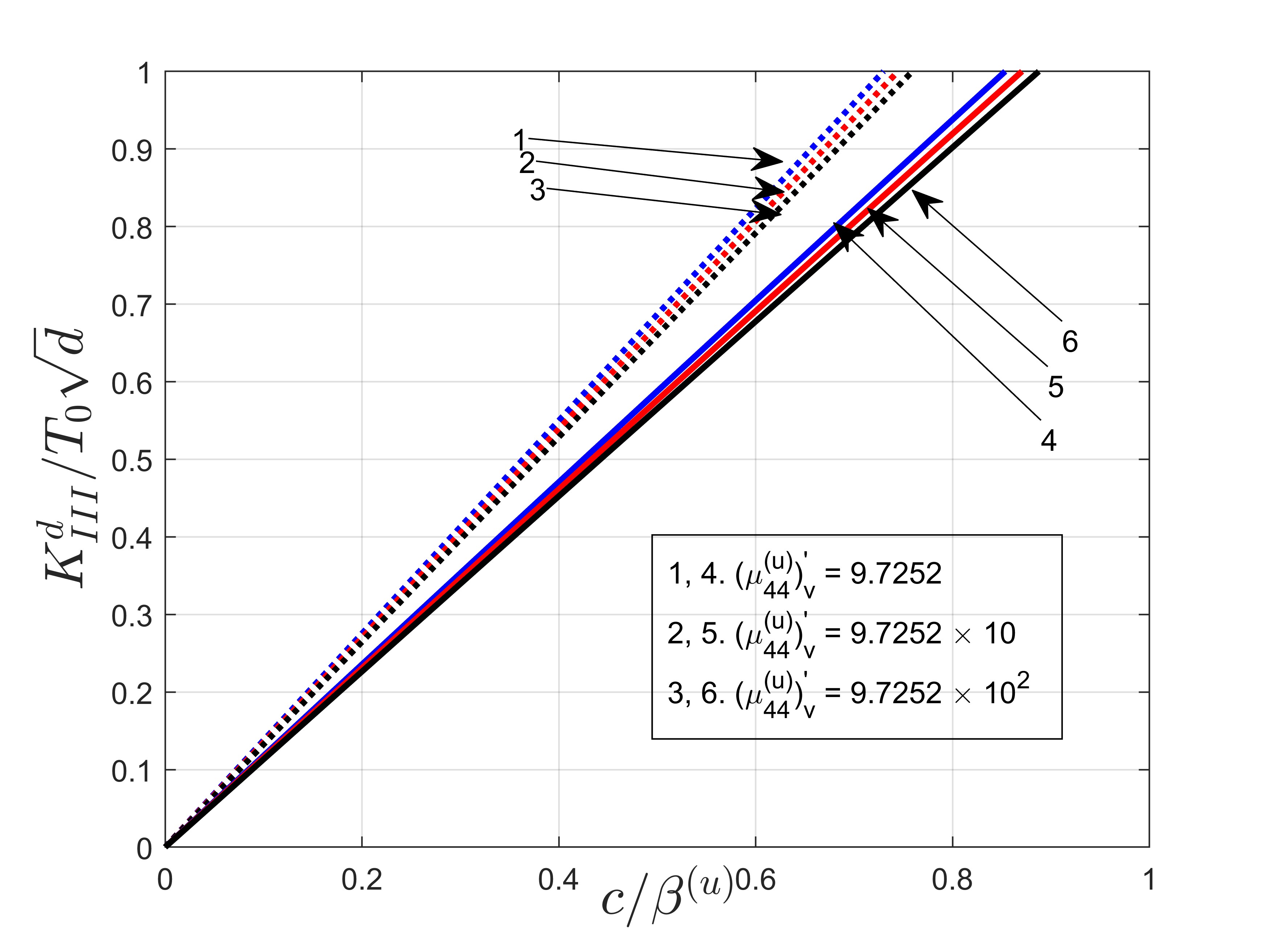}
\caption{}
\label{c44_u_v}
\end{subfigure}
~
\begin{subfigure}[b] {0.45\textwidth}
\includegraphics[width=\textwidth ]{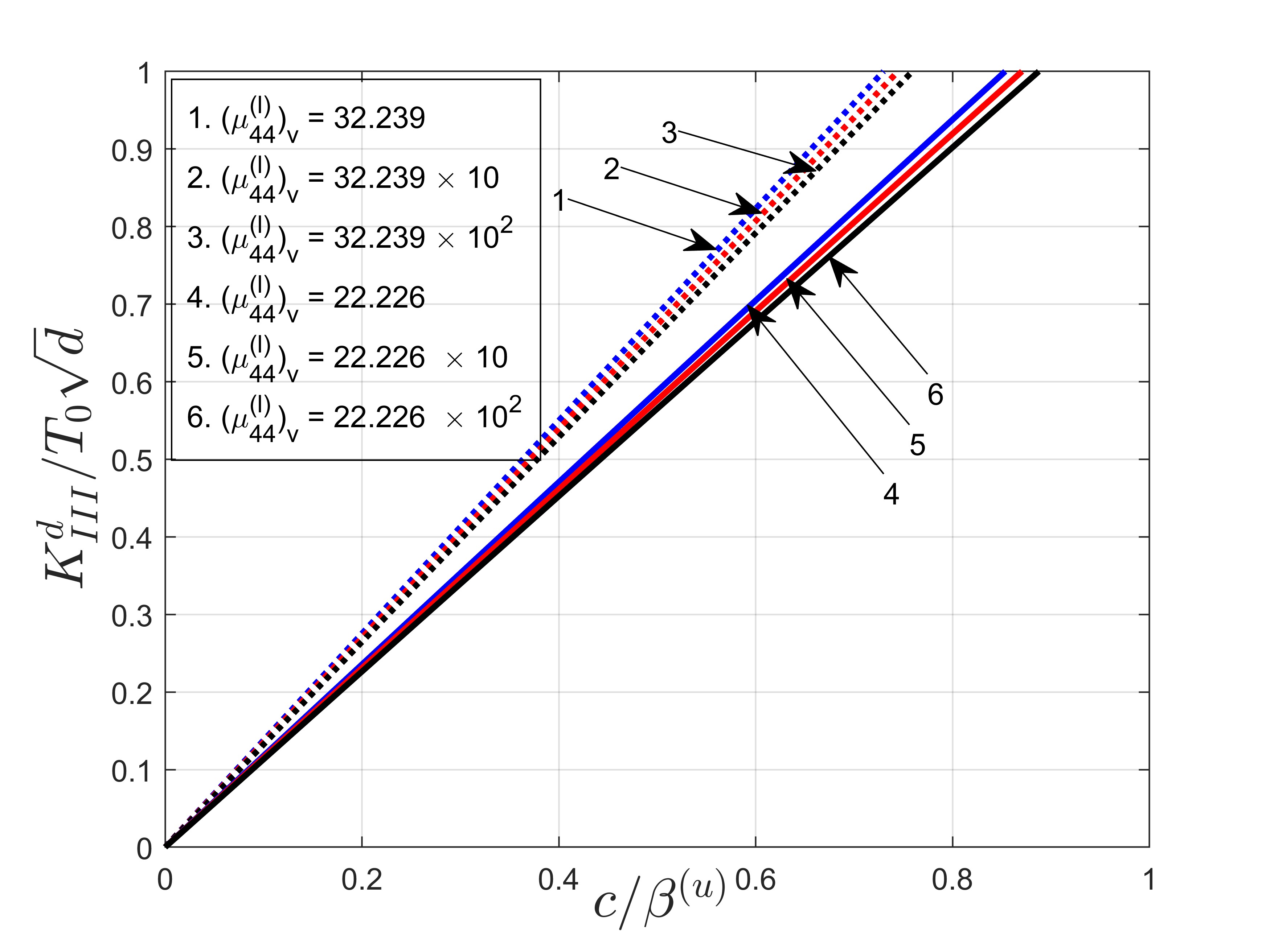}
\caption{}
\label{c44_l_v_edited}
\end{subfigure}

\caption{Variation of the SIF at the right crack tip as a function of the phase velocity of the Love wave due to variation in (a) heterogeneity parameter (b) spatial frequency (c) viscoelastic constant (upper-half) (d) viscoelastic constant (lower-half).}
\label{heterogenit_omega_visco}
\efg
%%%%%%%%%%%%%%%%%%%%%%%%%%%%%%%%%%%%%%
\bfg[htbp]
\centering
\begin{subfigure}[b] {0.45\textwidth}
\includegraphics[width=\textwidth ]{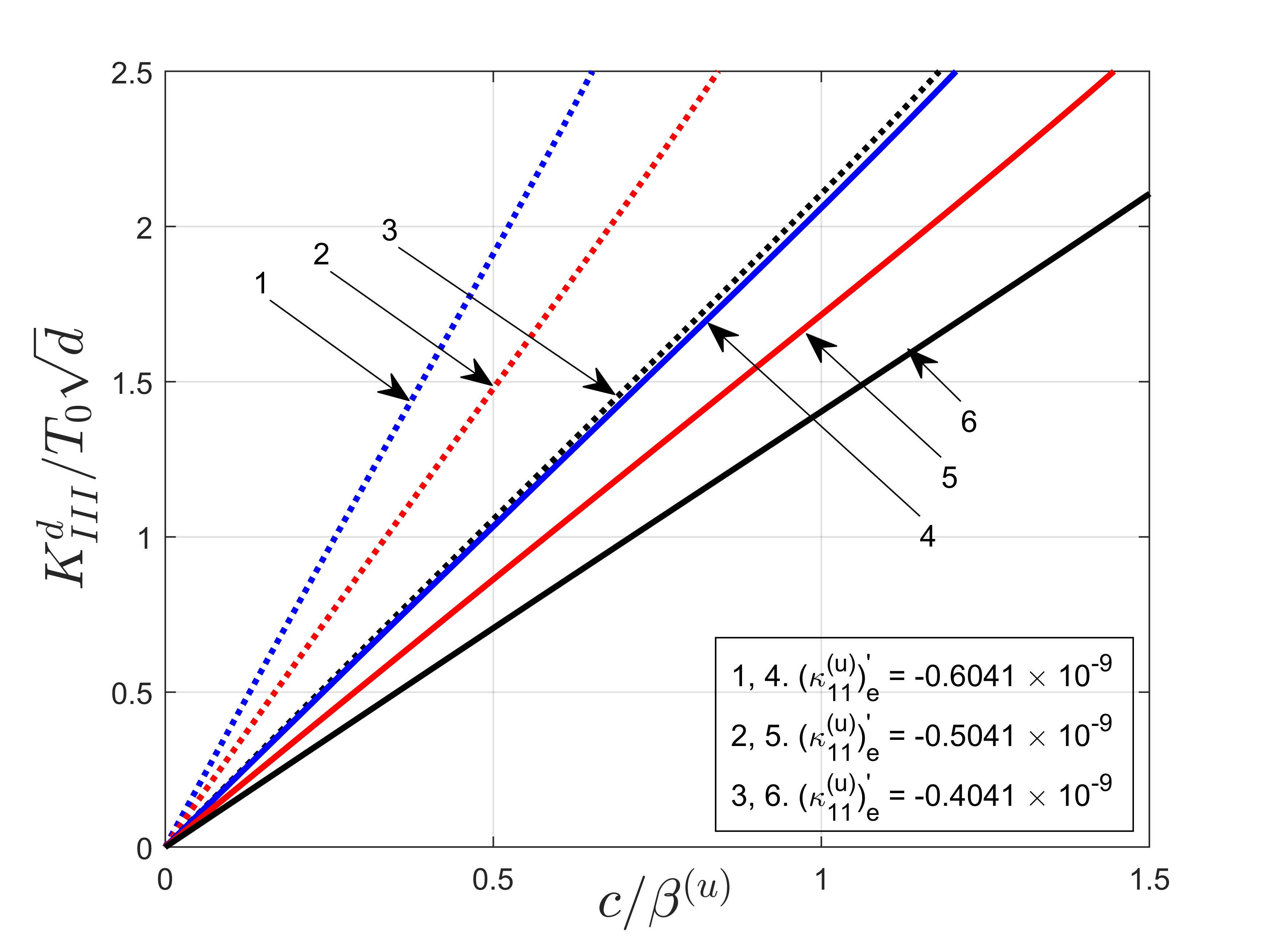}
\caption{}
\label{kappa1e}
\end{subfigure}
~
\begin{subfigure}[b] {0.45\textwidth}
\includegraphics[width=\textwidth ]{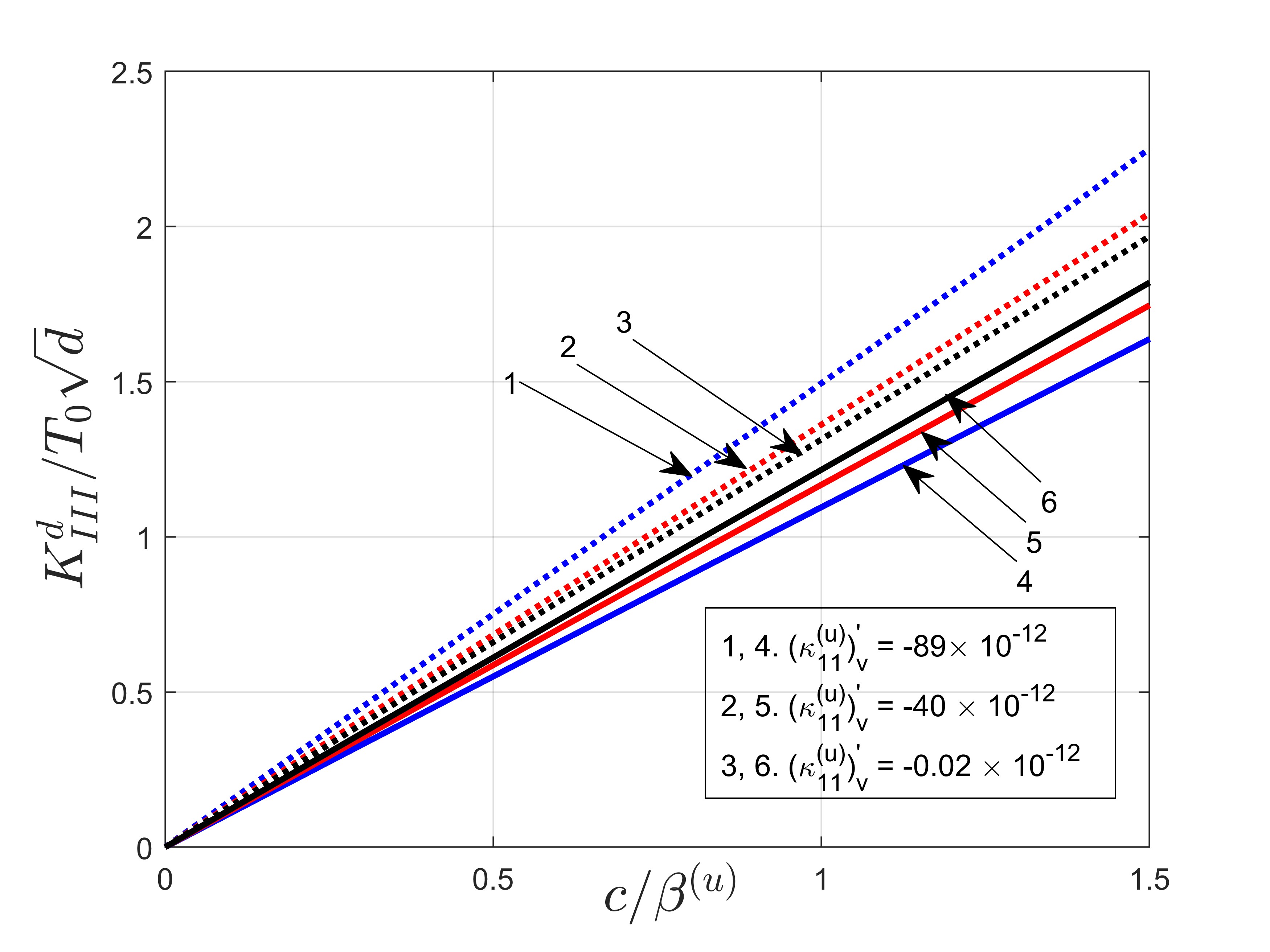}
\caption{}
\label{kappa1v}
\end{subfigure}
~
\begin{subfigure}[b] {0.45\textwidth}
\includegraphics[width=\textwidth ]{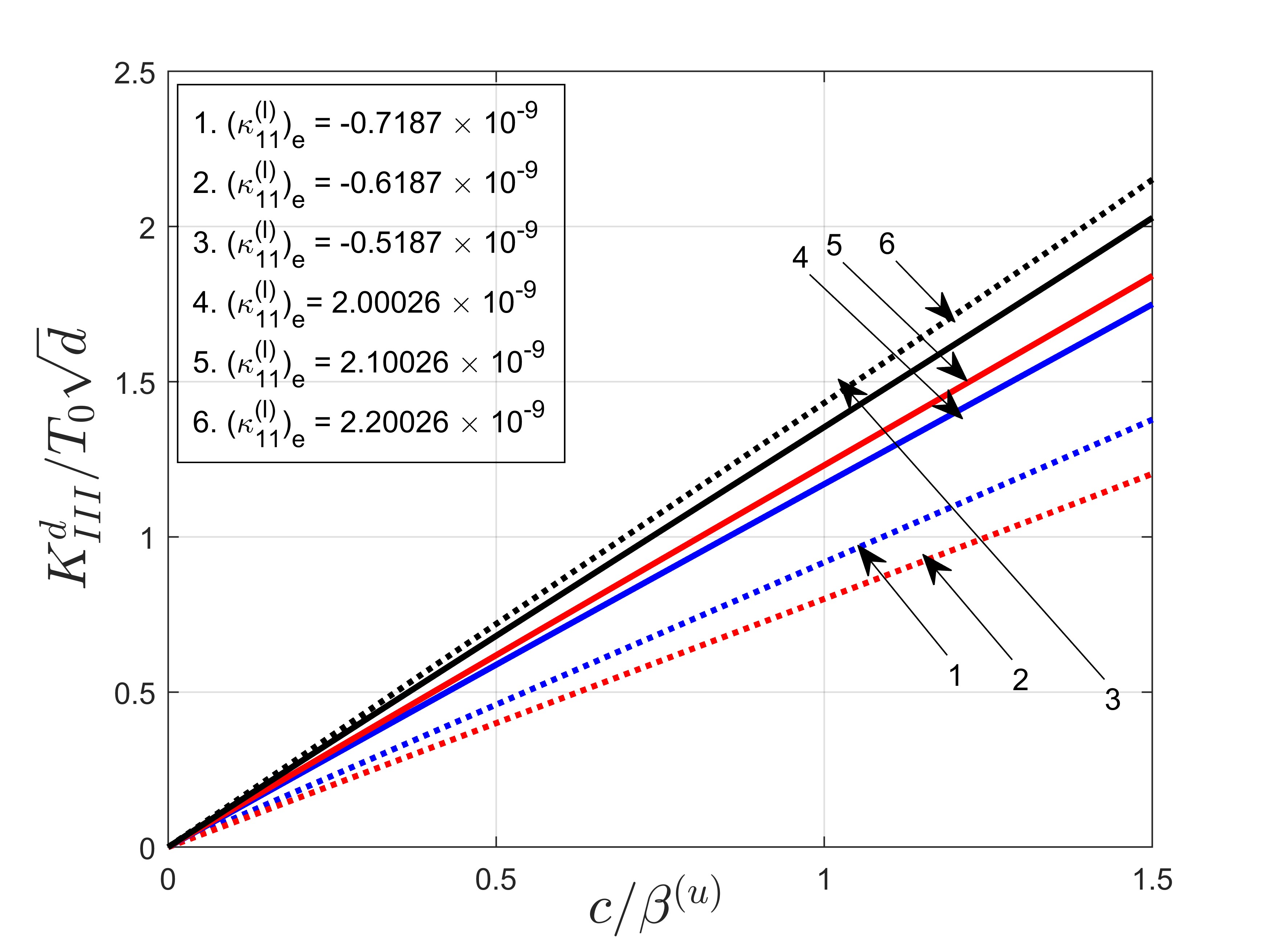}
\caption{}
\label{kappa2e}
\end{subfigure}
~
\begin{subfigure}[b] {0.45\textwidth}
\includegraphics[width=\textwidth ]{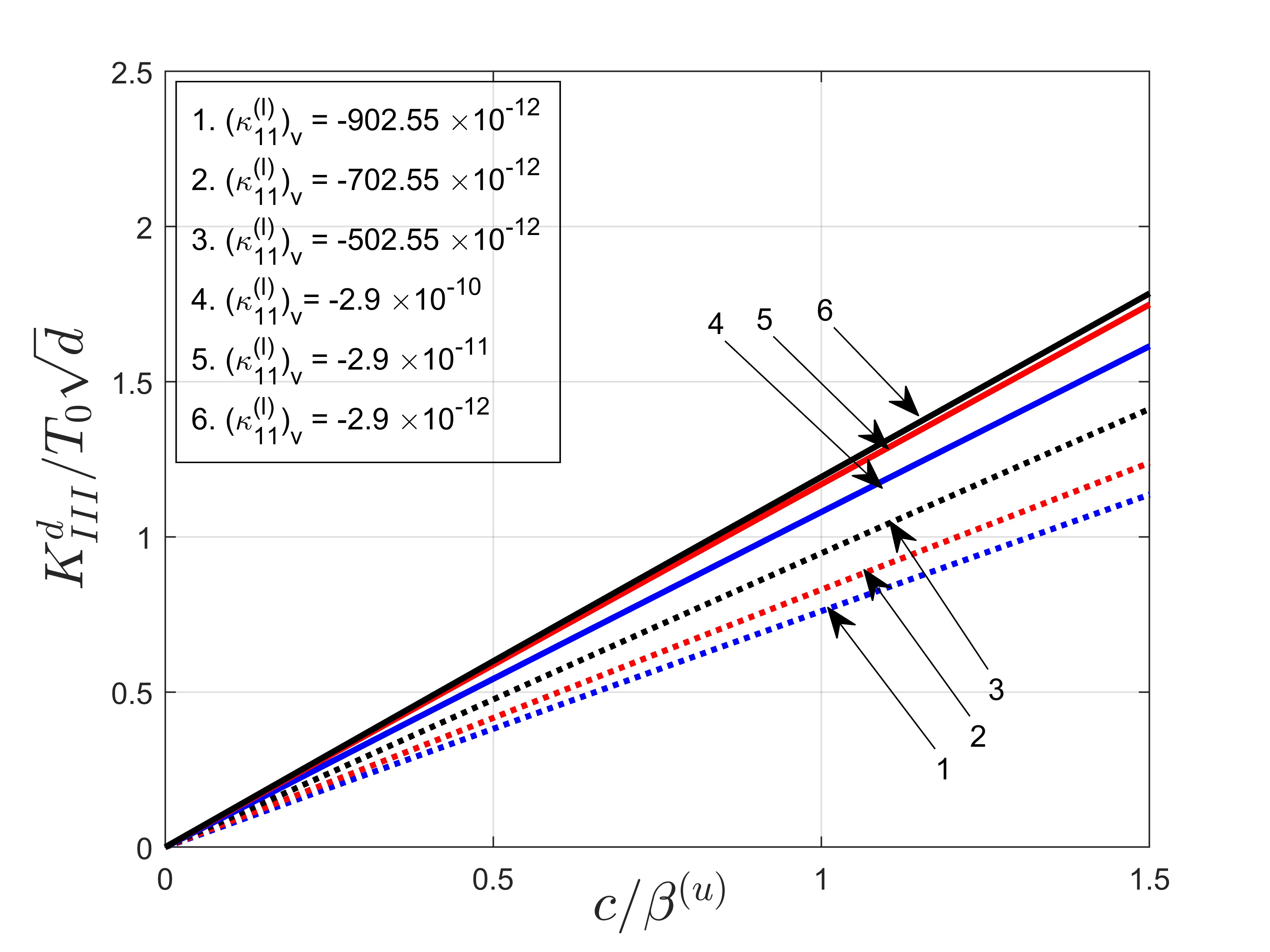}
\caption{}
\label{kappa2v}
\end{subfigure}

\caption{Variation of the SIF at the right crack tip as a function of the phase velocity of the Love wave due to variation in (a) dielectric constant (upper-half) (b) dielectric loss moduli (upper-half) (c) dielectric constant (lower-half) (d) dielectric loss moduli (lower-half).}
\label{variation_in_kappa}
\efg
\bfg[htbp]
\centering
\begin{subfigure}[b] {0.45\textwidth}
\includegraphics[width=\textwidth ]{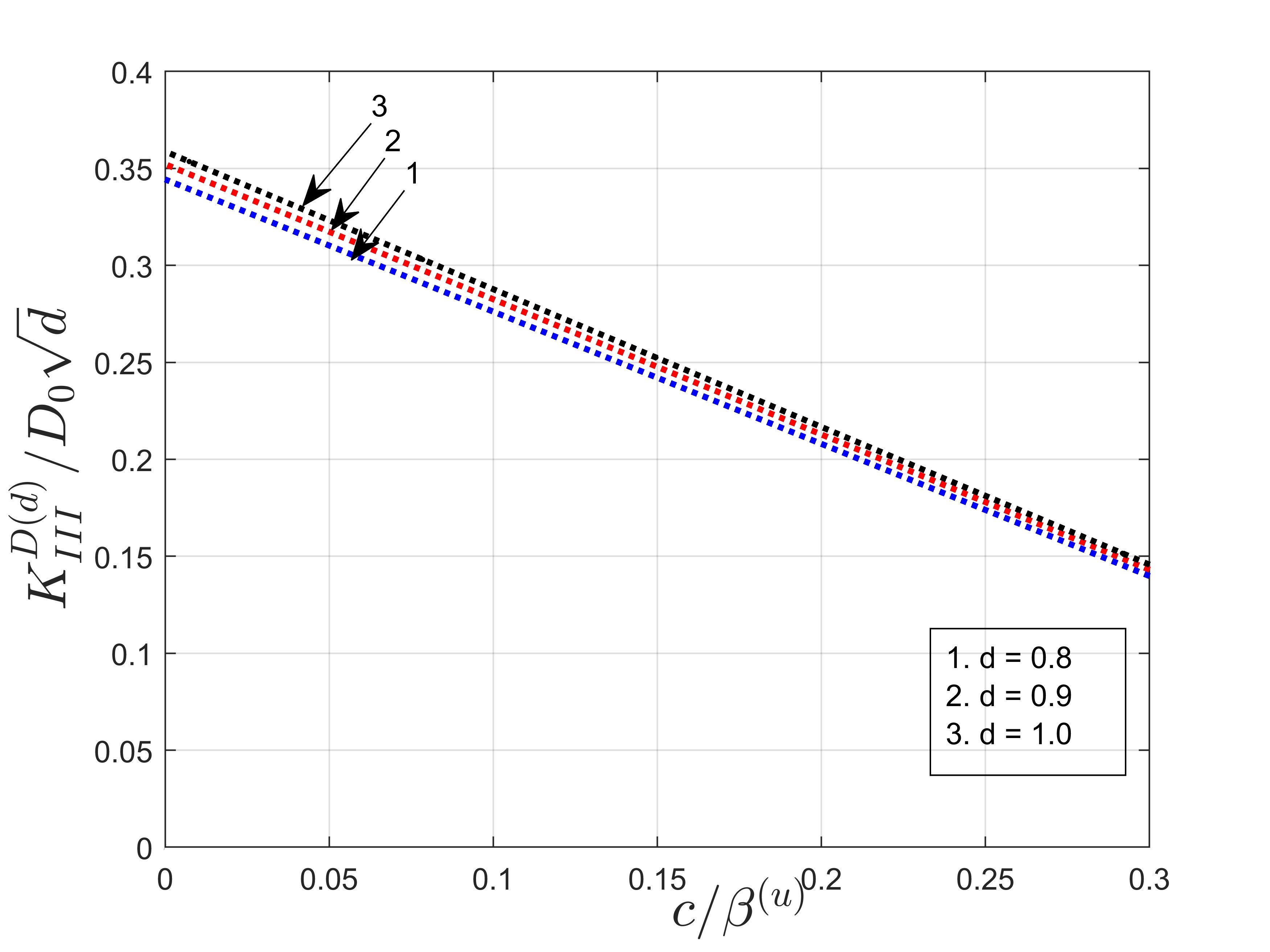}
\caption{}
\label{ED_d_BM1}
\end{subfigure}
~
\begin{subfigure}[b] {0.45\textwidth}
\includegraphics[width=\textwidth ]{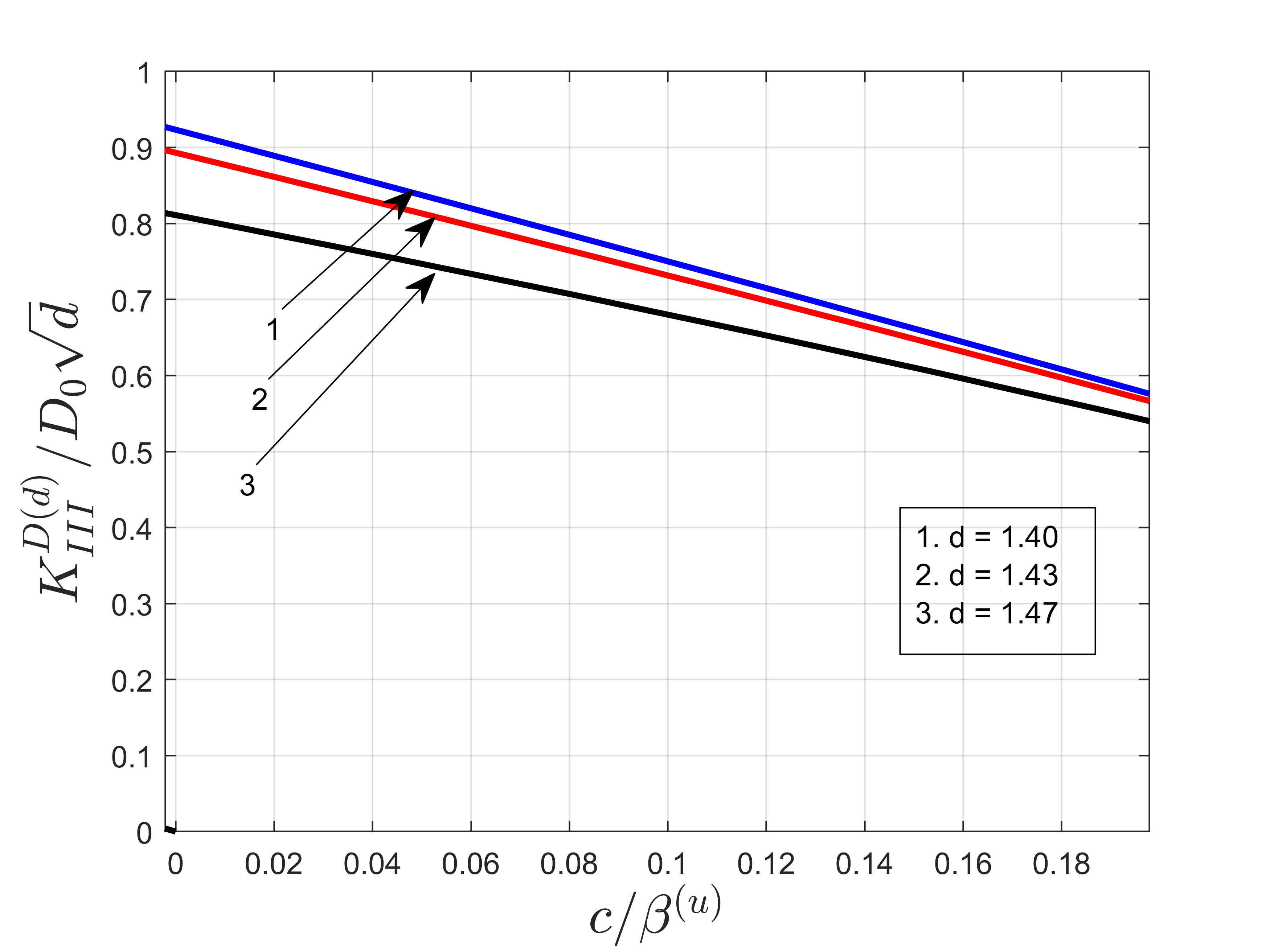}
\caption{}
\label{ED_d_BM2}
\end{subfigure}
~
\begin{subfigure}[b] {0.45\textwidth}
\includegraphics[width=\textwidth ]{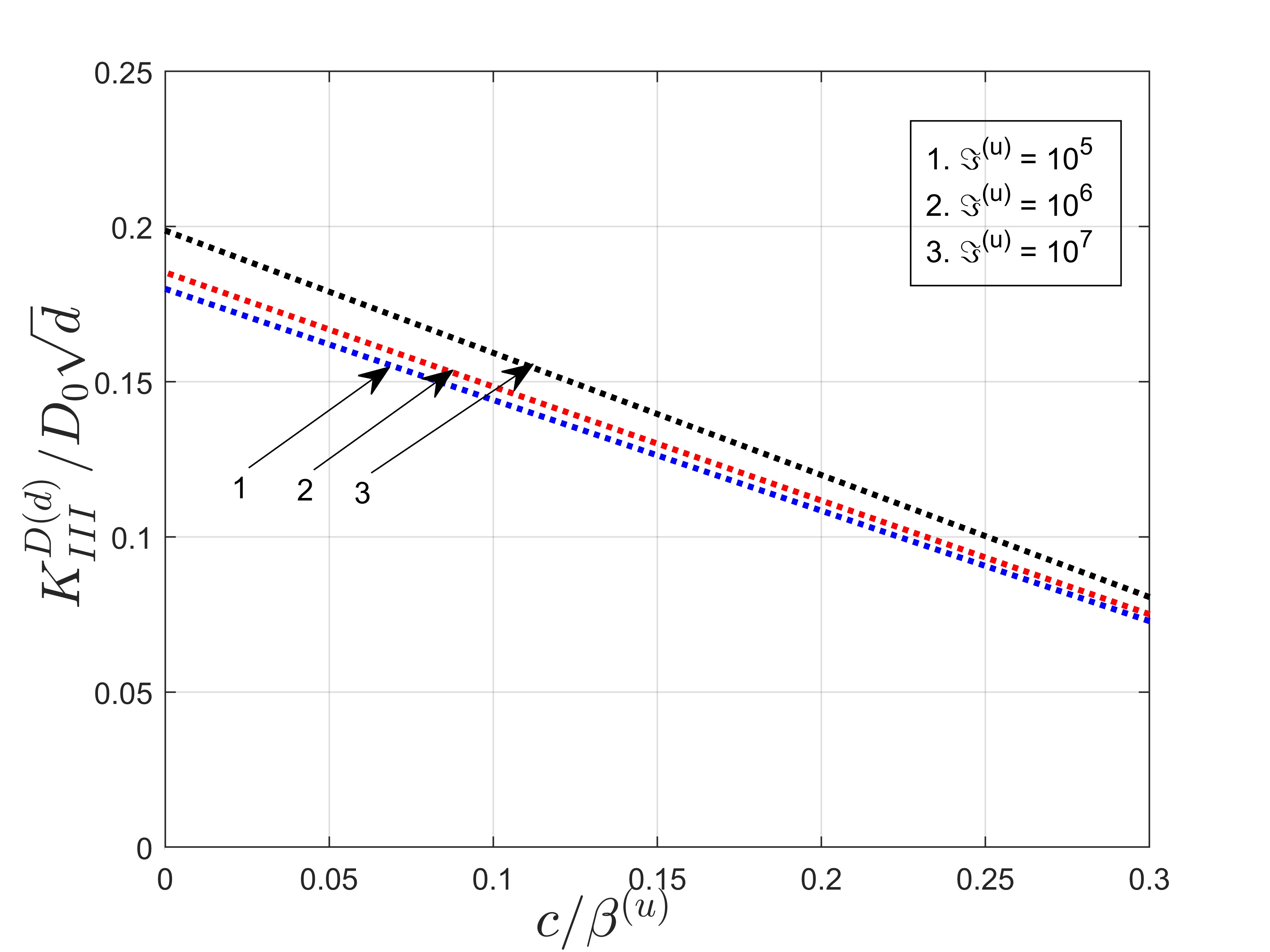}
\caption{}
\label{ED_IS_BM1}
\end{subfigure}
~
\begin{subfigure}[b] {0.45\textwidth}
\includegraphics[width=\textwidth ]{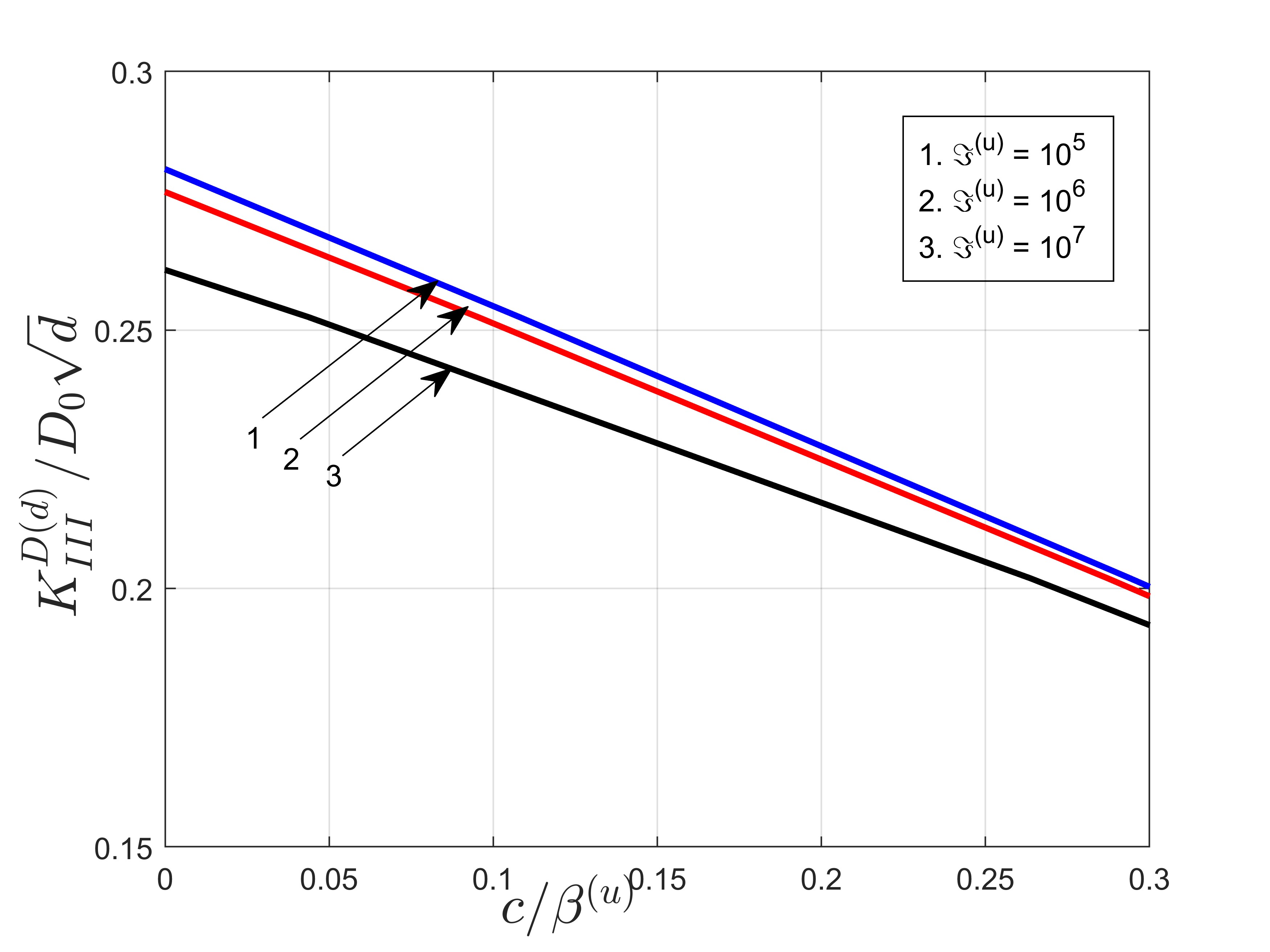}
\caption{}
\label{ED_IS_BM2}
\end{subfigure}

\caption{Variation of the EDIF at the right crack tip as a function of the phase velocity of the Love wave due to variation in (a) crack length (bimaterial I) (b) crack length (bimaterial II) (c) initial stress (bimaterial I) (d) initial stress (bimaterial II).}
\label{electric _displacement}
\efg
Figures \ref{variation in IS, d,D T} through \ref{variation_in_kappa} demonstrate that the stress SIF exhibits a direct proportionality with the phase velocity of the Love wave; specifically, as the phase velocity increases, the SIF also increases. 
Figure \ref{variation in IS, d,D T} represents the variation in SIF as a function of Love wave phase velocity with variation of initial stress (Fig. \ref{IS}), crack length (Fig. \ref{d}), interface stress (Fig. \ref{T0}), interface electric displacement (Fig. \ref{D0}), rotation parameter for upper (Fig. \ref{rotation1}) and lower half-spaces (Fig. \ref{rotation2}) respectively.
From Fig. \ref{IS}, it is evident that the SIF exhibits a differential response to increased pre-stress: it rises in bimaterial II but falls in bimaterial I. Furthermore, Figures \ref{d} and \ref{D0} depict a decrease in the SIF as both the crack length and the electric displacement within the crack at the interface increase. Conversely, Fig. \ref{T0} demonstrates that an increase in stress at the interface inside the crack results in a rise in the SIF for both bimaterial systems. Additionally, Fig. \ref{rotation1} illustrates that the SIF diminishes as the rotation parameter of the upper half-space increases. In contrast, Fig. \ref{rotation2} indicates that an increase in the rotation parameter of the lower half-space leads to an increase in the SIF for bimaterial I and a decrease for bimaterial II.

Figure \ref{vriation in e15} illustrates the variation in the SIF as a function of phase velocity with changes in the piezoelectric constant and piezoelectric loss moduli. It is observed from Fig. \ref{e15ue} that as the piezoelectric coefficient corresponding to the upper half-space increases, the SIF increases in both bimaterial systems. However, when varying the piezoelectric loss moduli corresponding to the upper half (see Fig. \ref{e15uv}) and the piezoelectric constant corresponding to the lower half (see Fig. \ref{e15le}), the SIF decreases in both bimaterial systems. Additionally, Fig. \ref{e15lv} shows that an increase in the piezoelectric loss moduli corresponding to the lower half leads to an increase in the SIF for bimaterial II, while it decreases for bimaterial I. Figure \ref{heterogenit_omega_visco} illustrates the variation in the SIF caused by the propagation of Love waves, influenced by changes in the heterogeneity parameter, spatial frequency, and viscoelastic parameter. From Fig. \ref{heterogenity_paramtre}, it is observed that as the heterogeneity parameter increases, the SIF decreases in both bimaterial systems. Conversely, SIF increases with the rise in the spatial frequency of the wave, as depicted in Fig. \ref{omegaedited}. Furthermore, Figs. \ref{c44_u_v} and \ref{c44_l_v_edited} demonstrate that an increase in the viscoelastic parameter leads to an increase in SIF across both bimaterial composite structures.

Figure \ref{variation_in_kappa} represents how dielectric constant and dielectric loss moduli will affect the SIF as a function of phase velocity. Figure \ref{kappa1e} shows that with an increase in the value of the dielectric constant corresponding to the upper half, the SIF decreases, whereas on increasing the value of dielectric loss moduli corresponding to the same half (see Fig. \ref{kappa1v}), SIF increases in case of bimaterial II and decreases in case of bimaterial I. From Figs. \ref{kappa2e} and \ref{kappa2v}, it is observed that the SIF increases in both bimaterial systems with higher dielectric constant and dielectric loss moduli of the lower half-space. Figure \ref{electric _displacement} illustrates the relationship between the electric displacement intensity factor (EDIF) and the Love wave phase velocity, considering different crack lengths and initial stress levels. It is observed that the EDIF decreases as the phase velocity of the Love wave increases. This behavior indicates that the energy associated with the electric displacement intensity diminishes as the wave propagates faster. Figures \ref{ED_d_BM1} and \ref{ED_d_BM2} represent the variation in the EDIF due to changes in crack length. It is observed that with an increase in crack length, the EDIF increases in the case of Bimaterial I, whereas it decreases in the case of Bimaterial II. Additionally, Figs. \ref{ED_IS_BM1} and \ref{ED_IS_BM2}, which are plotted for variations in initial stress, show that with an increase in initial stress, the EDIF increases in Bimaterial I and decreases in Bimaterial II.

% \section{Special cases} 
% \label{special cases}
Figure \ref{special_case1} illustrates a special case showing the impact of pre-stress and rotation parameters on the SIF for the two bimaterial models considered. Figures \ref{special case 3 BM1} and \ref{special case 3 BM2} have been plotted for bimaterial I (Epoxy-BNKLBT and Epoxy-KNLNTS) and bimaterial II (Epoxy-BNKLBT and Epoxy-PZT7A), respectively. These observations reveal that both initial stress and rotational parameters lead to an increase in the SIF in the vicinity of the crack tip. Furthermore, the SIF is observed to increase more significantly in bimaterial II compared to bimaterial I, reflecting a greater sensitivity of the SIF to these parameters in the second material.
%%%%%%%%%%%%%%%%%%%%%%%%%%%%%%%%%%%%%%%%%%%%%%%%%%%%%%%%%%%%
 \bfg[htbp]
\centering
\begin{subfigure}[b] {0.45\textwidth}
\includegraphics[width=\textwidth ]{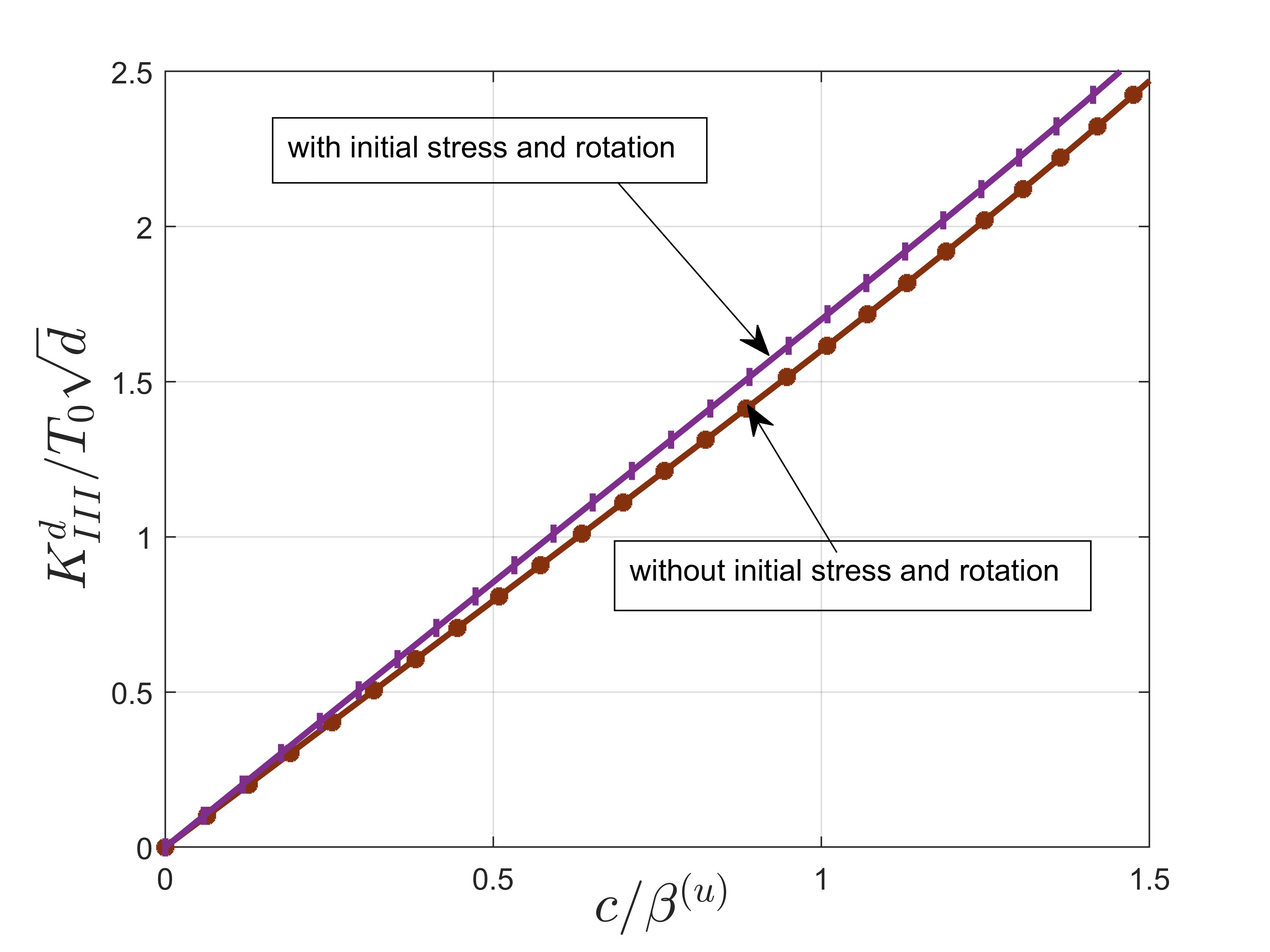}
\caption{}
\label{special case 3 BM1}
\end{subfigure}
~
\begin{subfigure}[b] {0.45\textwidth}
\includegraphics[width=\textwidth ]{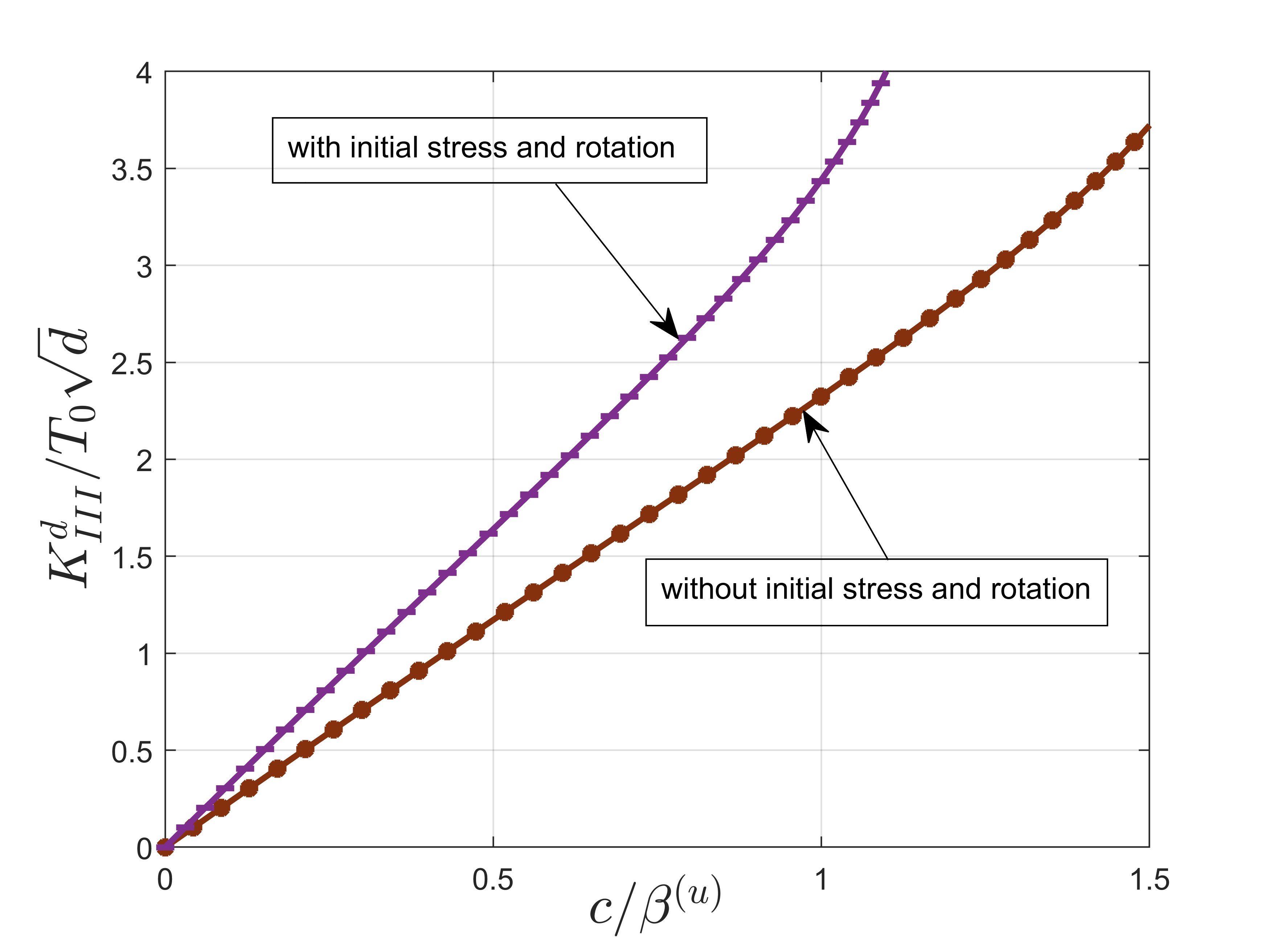}
\caption{}
\label{special case 3 BM2}
\end{subfigure}

\caption{Effect of initial stress and rotation on dimensionless SIF corresponding to (i) bimaterial I ( Epoxy-BNKLBT and Epoxy-KNLNTS) (ii) bimaterial II (Epoxy-BNKLBT and Epoxy-PZT7A). }
\label{special_case1}
\efg

\section{Conclusion}
\label{conclusion}
A novel aspect of this study is the analysis of dynamic mode-III fracture interaction with the propagating Love wave in a pre-stressed FGPVM half-space bonded to PVM half-space under rotation. The significant impacts of physical characteristics, such as initial stress, crack length, rotation parameter, interface stress, interface electric displacement, and velocity of Love wave on SIF and EDIF have been examined. The key observations from the obtained results are: 
\begin{itemize}
\item In the dynamic mode-III fracture scenario within a pre-stressed piezoelectric viscoelastic composite, the SIF increases with the rising Love wave velocity in the analyzed structure.
\item As Love wave phase velocity increases, the EDIF decreases, indicating a reduction in the energy linked with the electric displacement as the wave propagates faster.
\item As the piezoelectric constant of the upper half-space increases, the SIF in both bimaterial structures also increases due to enhanced coupling between the electric field and mechanical stress, resulting in greater stress accumulation at the crack tip. However, when the piezoelectric loss modulus of the same half-space increases, the SIF decreases as the higher energy dissipation reduces the effective stress concentration at the crack tip.
\item As the heterogeneity factor increases in the bimaterial system, the SIF decreases in both the considered bimaterials, contributing to the engineering structure's longevity. This reduction in SIF is significant because a lower SIF indicates a diminished likelihood of crack propagation, thereby enhancing the durability and service life of the structure.
\item Increases in initial stresses and the rotational parameter lead to higher SIF in both bimaterials, underscoring their significant influence on stress concentration around cracks.
\item With increasing crack length, the SIF decreases for both considered bimaterial systems. In contrast, the EDIF increases in bimaterial I but decreases in bimaterial II.
\end{itemize}
The present study is useful not only for advancing technologies in cancer detection and biomedical implants but also has significant implications for the aerospace industry, supporting the adoption of damage tolerance analysis to enhance the safety and reliability of aircraft and spacecraft.

\noindent \textbf{Acknowledgements}\\
\noindent The authors sincerely thank the National Institute of Technology Hamirpur (Department of Mathematics and Scientific Computing), Hamirpur (India), for providing the research fellowship for a PhD to Ms. Diksha.

\noindent \textbf{Compliance with ethical standards}\\
\noindent Conflict of interest: The authors declare that they have no conflict of interest.

\bibliographystyle{unsrt}
\bibliography{reference}
\end{document}